\documentclass[12pt]{article}
\usepackage{a4p}
\usepackage{epsfig}
\usepackage{array}
\usepackage{cite}
\usepackage{pennames}
\usepackage{rotating}
\newcommand{\PPEnum} {CERN-EP/99-097}

\newcommand{\Date}      {8th July 1999}
\newcommand{\inmath}[1] {\ifmmode#1\else$#1$\fi}
\newcommand{\definmath}[2] {\def#1{\ifmmode#2\else$#2$\fi}}
\newcommand{\alphas} {\alpha_{\mathrm{s}}}
\newcommand{\alphaem} {\mbox{$\alpha_{\mathrm{em}}$}}

\definmath{\PWpm} {\mathrm{W}^{\pm}}      
\definmath{\Plp} {\ell^{+}}        
\definmath{\Plm} {\ell^{-}}        
\definmath{\Plpm}   {\ell^{\pm}}         
\definmath{\Pgtp} {\tau^{+}}        
\definmath{\Pgtm} {\tau^{-}}        
\definmath{\Pgtpm}   {\tau^{\pm}}         
\definmath{\Pgn}  {\nu}          
\definmath{\Pagn} {\overline{\nu}}     
\definmath{\Pf}      {\mathrm{f}}
\definmath{\Paf}  {\overline{\mathrm{f}}}
\definmath{\Pq}      {\mathrm{q}}
\definmath{\Paq}  {\overline{\mathrm{q}}}
\definmath{\Pu}      {\mathrm{u}}
\definmath{\Pau}  {\overline{\mathrm{u}}}
\definmath{\Pd}      {\mathrm{d}}
\definmath{\Pad}  {\overline{\mathrm{d}}}
\definmath{\Ps}      {\mathrm{s}}
\definmath{\Pas}  {\overline{\mathrm{s}}}
\definmath{\Pc}      {\mathrm{c}}
\definmath{\Pac}  {\overline{\mathrm{c}}}
\definmath{\Pb}      {\mathrm{b}}
\definmath{\Pab}  {\overline{\mathrm{b}}}
\definmath{\Pt}      {\mathrm{t}}
\definmath{\Pat}  {\overline{\mathrm{t}}}
\definmath{\Pap}  {\overline{\mathrm{p}}}
\definmath{\Pan}  {\overline{\mathrm{n}}}
\definmath{\PaD}  {\overline{\mathrm{D}}}
\definmath{\PaDz} {\overline{\mathrm{D}}^{0}}
\definmath{\PaB}  {\overline{\mathrm{B}}}
\definmath{\PaBz} {\overline{\mathrm{B}}^{0}}
\definmath{\PsDpm}   {\mathrm{D}^{\pm}_{\mathrm{s}}}  
\definmath{\PcgLpm}  {\Lambda^{\pm}_{\mathrm{c}}}  
\definmath{\PD} {\mathrm{D}}     
\definmath{\PDst} {\mathrm{D}^{*}}     
\definmath{\PgLz} {\Lambda^{0}}        

\newcommand{\snu}{\tilde{\nu}}

\newcommand{\massof}[1] {m_{\smash{#1}\mathstrut}}
\newcommand{\mtop}   {\massof{\mathrm{top}}}
\newcommand{\mHiggs} {\massof{\mathrm{Higgs}}}

\newcommand{\mPZ} {\massof{\mathrm{Z}}}
\newcommand{\mX}  {\massof{\mathrm{X}}}
\newcommand{\AFB}    {A_{\mathrm{FB}}}
\newcommand{\AFBSM}  {A_{\mathrm{FB}}^{\mathrm{SM}}}

\newcommand{\thacol}{\theta_{\mathrm{acol}}}

\newcommand{\epem}   {\Pep\Pem}

\newcommand{\mumu}   {\Pgmp\Pgmm}
\newcommand{\tautau} {\Pgtp\Pgtm}
\newcommand{\ffbar}  {\Pf\Paf}

\newcommand{\qqbar}  {\Pq\Paq}
\newcommand{\uubar}  {\Pu\Pau}
\newcommand{\ddbar}  {\Pd\Pad}

\newcommand{\ud}     {\uubar + \ddbar}

\newcommand{\WW}{\ensuremath{\mathrm{W}^+\mathrm{W}^-}}
\newcommand{\eetoee}    {\epem\to\epem}

\newcommand{\eetomumu}     {\epem\to\mumu}
\newcommand{\eetotautau}   {\epem\to\tautau}

\newcommand{\eetoqq}    {\epem\to\qqbar}

\newcommand{\dsdcc}        {{\rm d}\sigma/{\rm d}\!\cos\theta}
\newcommand{\dsdabscc}     {{\rm d}\sigma/{\rm d}|\cos\theta|}
\def\lapproxeq {\mbox{{\lower .7ex\hbox{$\;\stackrel{\textstyle
                  <}{\sim}\;$}}}}
\def\gapproxeq  {\mbox{{\lower .7ex\hbox{$\;\stackrel{\textstyle
                  >}{\sim}\;$}}}}
\newcommand{\roots} {\sqrt{s}}

\newcommand {\ct}      {\mbox{$\cos \theta$}}
\newcommand {\absct}   {\mbox{$|\cos \theta |$}}

\newcommand {\absctem} {\mbox{$|\cos \theta_{{\rm e}^-} |$}}

\newcommand {\absctep} {\mbox{$|\cos \theta_{{\rm e}^+} |$}}
\newcommand {\absctepem} {\mbox{$|\cos \theta_{{\rm e}^\pm} |$}}

\definmath{\GeV}  {\mathrm{GeV}}
\definmath{\GeVc} {\mathrm{GeV}\!/c}
\definmath{\GeVcc}   {\mathrm{GeV}\!/c^2}
\definmath{\MeV}  {\mathrm{MeV}}
\definmath{\MeVc} {\mathrm{MeV}\!/c}
\definmath{\MeVcc}   {\mathrm{MeV}\!/c^2}
\definmath{\MVm}  {\mathrm{MV}\!/\mathrm{m}}
\definmath{\keV}  {\mathrm{keV}}
\definmath{\keVcm}   {\mathrm{keV}\!/\mathrm{cm}}
\definmath{\kV}      {\mathrm{kV}}
\definmath{\km}      {\mathrm{km}}
\definmath{\meter}   {\mathrm{m}}
\definmath{\cm}      {\mathrm{cm}}
\definmath{\mm}      {\mathrm{mm}}
\definmath{\micron}  {\mu\mathrm{m}}
\definmath{\nm}      {\mathrm{nm}}
\definmath{\kg}      {\mathrm{kg}}
\definmath{\gram} {\mathrm{g}}
\definmath{\second}  {\mathrm{s}}
\definmath{\microsec}   {\mu\mathrm{s}}
\definmath{\degree}  {^\circ}
\definmath{\degC} {^\circ\mathrm{C}}
\definmath{\ohm}  {\Omega}
\definmath{\Mohm} {\mathrm{M}\Omega}
\definmath{\rad}  {\mathrm{rad}}
\definmath{\mrad} {\mathrm{mrad}}
\definmath{\nb}      {\mathrm{nb}}
\newcommand{\eqref}[1]  {(\ref{#1})}
\newcommand{\PhysLett}  {Phys.~Lett.}
\newcommand{\PRL} {Phys.~Rev.\ Lett.}
\newcommand{\PhysRep}   {Phys.~Rep.}
\newcommand{\PhysRev}   {Phys.~Rev.}
\newcommand{\NPhys}  {Nucl.~Phys.}
\newcommand{\NIM} {Nucl.~Instr.\ and Meth.}
\newcommand{\ZPhys}  {Z.~Phys.}
\newcommand{\IEEENS} {IEEE Trans.\ Nucl.~Sci.}
\newcommand{\CPC} {Comp. Phys. Comm.}
\newcommand{\EPJ} {Eur.~Phys.~J.} 
\newcommand{\ALEPHColl}   {ALEPH Collab.}
\newcommand{\DELPHIColl}  {DELPHI Collab.}

\newcommand{\LthreeColl}  {L3 Collab.}
\newcommand{\OPALColl}    {OPAL Collab.}
\newcolumntype{L} {>{$}l<{$}}
\newcolumntype{C} {>{$}c<{$}}
\newcolumntype{R} {>{$}r<{$}}

   \newcommand{\lept}
{\ell^+\ell^-}   

\newcommand{\epsz}  {\varepsilon_0}   \newcommand{\lamm}   {\Lambda_-}
\newcommand{\lamp} {\Lambda_+} 

\begin{document}
%
%
\begin{titlepage}
%
\begin{center}
    \Large EUROPEAN LABORATORY FOR PARTICLE PHYSICS
\end{center}
\bigskip 
\begin{flushright}
    \large \PPEnum \\  \Date
\end{flushright}
%
%
\begin{center}
    \huge\bf\boldmath Tests of the Standard Model and Constraints on
    New Physics from Measurements of Fermion-pair Production at
    189~\GeV\ at LEP 
\end{center}
\vspace{0.5cm} 
%
%
\begin{center}
    \LARGE   The  OPAL   Collaboration   \\  
\vspace{0.5cm} 
%
%
\begin{abstract}
Cross-sections and angular distributions for hadronic and lepton pair final
states in \epem\ collisions at a centre-of-mass energy near 189~GeV, 
measured with the OPAL detector at LEP, are presented and compared with the 
predictions of the Standard Model. 
The results are used to measure the energy 
dependence of the electromagnetic coupling constant \alphaem, and to place 
limits on new physics as described by four-fermion 
contact interactions or by the exchange of a new heavy particle such as a 
sneutrino in supersymmetric theories with 
$R$-parity violation. A search for the indirect effects of the gravitational
interaction in extra dimensions on the \mumu\ and \tautau\ final states
is also presented. 
\end{abstract}
\end{center}
\vspace{0.5cm}
\begin{center}
{\large Submitted to European Journal of Physics C}
\end{center}
%
%
\end{titlepage}
\begin{center}{\Large        The OPAL Collaboration
}\end{center}\bigskip
\begin{center}{
G.\thinspace Abbiendi$^{  2}$,
K.\thinspace Ackerstaff$^{  8}$,
G.\thinspace Alexander$^{ 23}$,
J.\thinspace Allison$^{ 16}$,
K.J.\thinspace Anderson$^{  9}$,
S.\thinspace Anderson$^{ 12}$,
S.\thinspace Arcelli$^{ 17}$,
S.\thinspace Asai$^{ 24}$,
S.F.\thinspace Ashby$^{  1}$,
D.\thinspace Axen$^{ 29}$
% ...
}\end{center}\bigskip
\bigskip
$^{  1}$School of Physics and Astronomy, University of Birmingham,
Birmingham B15 2TT, UK
\newline
$^{  2}$Dipartimento di Fisica dell' Universit\`a di Bologna and INFN,
I-40126 Bologna, Italy
\newline
% ...
$^{  8}$CERN, European Organisation for Particle Physics,
CH-1211 Geneva 23, Switzerland
\newline
$^{  9}$Enrico Fermi Institute and Department of Physics,
University of Chicago, Chicago IL 60637, USA
\newline
% ...
$^{ 12}$Indiana University, Department of Physics,
Swain Hall West 117, Bloomington IN 47405, USA
\newline
% ...
$^{ 16}$Department of Physics, Schuster Laboratory, The University,
Manchester M13 9PL, UK
\newline
$^{ 17}$Department of Physics, University of Maryland,
College Park, MD 20742, USA
\newline
% ...
$^{ 23}$Department of Physics and Astronomy, Tel Aviv University,
Tel Aviv 69978, Israel
\newline
$^{ 24}$International Centre for Elementary Particle Physics and
Department of Physics, University of Tokyo, Tokyo 113-0033, and
Kobe University, Kobe 657-8501, Japan
\newline
% ...
$^{ 29}$University of British Columbia, Department of Physics,
Vancouver BC V6T 1Z1, Canada
\newline
% ...
 Debrecen, Hungary.
\clearpage

 
\section{Introduction}           \label{sec:intro}
Measurements of fermion-pair production in \epem\ collisions at high
energies provide an important test of Standard Model predictions, and
allow limits to be set on many possible new physics 
processes~\cite{bib:OPAL-SM172,bib:OPAL-SM183,bib:ADL-SM}.
In this paper we present measurements of cross-sections and angular
distributions for hadronic and lepton pair final states at a centre-of-mass 
energy $\sqrt{s}$ near 189~GeV; forward-backward asymmetries for the 
leptonic states are also given. The data were collected by the OPAL detector 
at LEP in 1998. 

The analyses presented here are essentially the same as those already
presented at lower energies~\cite{bib:OPAL-SM183,bib:OPAL-SM172}. We use 
identical techniques to measure $s'$, the square of the centre-of-mass
energy of the \epem\ system after initial-state radiation, and 
to separate `non-radiative' events, which have little initial-state
radiation, from `radiative return' to the Z peak. As at 
183~GeV~\cite{bib:OPAL-SM183}, we define non-radiative events as those
having $s'/s > 0.7225$, and inclusive measurements are corrected to 
$s'/s > 0.01$. We correct our measurements of hadronic, \mumu\ and \tautau\ 
events for the effect of interference between initial- and final-state 
radiation, as in our previous publications, and also use the same treatment 
of the four-fermion contribution to the two-fermion final states. Because of 
ambiguities arising from the $t$-channel contribution, for the \epem\ final 
state the acceptance is defined in terms of the angle $\theta$ of the 
electron or positron with respect to the electron beam direction and the 
acollinearity angle $\thacol$ between the electron and positron. Cross-sections
and asymmetries for \epem\ are not corrected for interference between initial-
and final-state radiation; they are compared to theoretical predictions
which include interference. With the higher luminosity and hence higher
statistics available at 189~\GeV\ we have been able to reduce the
experimental systematic errors in some channels, compared with previous
analyses.

Measurements of fermion-pair production up to 183~GeV have shown very 
good agreement with Standard Model
expectations~\cite{bib:OPAL-SM172,bib:OPAL-SM183,bib:ADL-SM}. Here we repeat
our measurement of the electromagnetic coupling constant \alphaem($\sqrt{s}$)
including the higher energy data. Including data at 189~GeV also
allows us to extend the searches for new physics presented 
in~\cite{bib:OPAL-SM183}. In particular we obtain improved limits
on the energy scale of a possible four-fermion contact interaction.
We also present results of a search for particles which couple to leptons,
such as scalar neutrinos (sneutrinos) in theories with $R$-parity violation.
These analyses are updates of those already presented 
in~\cite{bib:OPAL-SM183}. Recently it has been pointed out that the
quantum-gravity scale could be as low as the electroweak scale with
gravitons propagating in extra dimensions~\cite{bib:ADD}. Indirect
effects of such gravitational interactions might be seen at 
colliders~\cite{bib:Giudice}. In this paper we present a new search for
the possible effects of the gravitational interaction in extra dimensions
on the \mumu\ and \tautau\ final states. We have obtained lower
limits on the effective Planck scale in the space with extra dimensions.

The paper is organized as follows. In Section~\ref{sec:data} we describe
the data analysis, cross-section and asymmetry measurements. Since the
analyses are essentially the same as in~\cite{bib:OPAL-SM172,bib:OPAL-SM183} 
we give only a brief description of any changes. In the earlier
analyses the errors were generally dominated by statistics, but with the
much larger data sample available at 189~GeV the systematic errors
are now often comparable with the statistical ones. We therefore
discuss the estimation of systematic errors in some detail. 
In Section~\ref{sec:sm} we
compare our measurements to the predictions of the Standard Model
and use them to measure the energy dependence of $\alphaem$. The results of 
searches for new physics are presented in Section~\ref{sec:new_phys}.

\section{Data Analysis}           \label{sec:data}
The OPAL detector\footnote{OPAL uses a right-handed coordinate system in
which the $z$ axis is along the electron beam direction and the $x$
axis is horizontal. The polar angle $\theta$ is measured with respect
to the $z$ axis and the azimuthal angle $\phi$ with respect to the
$x$ axis.}, trigger and data acquisition system are fully described 
elsewhere~\cite{bib:OPAL-detector,bib:OPAL-SI,bib:OPAL-SW,bib:OPAL-TR,
bib:OPAL-DAQ}. The high redundancy of the trigger system leads to
negligible trigger inefficiency for all channels discussed here.
The analyses presented in this paper use more than
180~pb$^{-1}$ of data collected at centre-of-mass energies near 189~GeV 
during 1998; the actual amount of data varies from channel to channel. The
luminosity-weighted mean centre-of-mass energy is
188.63$\pm$0.04~GeV~\cite{bib:ELEP}. 

Selection efficiencies and backgrounds were calculated using Monte
Carlo simulations. The default set of generators used is identical to that
in~\cite{bib:OPAL-SM183}. Use of alternative generators in assessing
sytematic errors is discussed below.
All events were passed through a full simulation~\cite{bib:gopal} of the OPAL 
detector and processed as for real data.

The luminosity was measured using small-angle Bhabha scattering events
recorded in the silicon-tungsten luminometer~\cite{bib:OPAL-SW,bib:OPAL-SM172}.
The overall error on the luminosity measurement amounts to 0.21\%,
arising from data statistics (0.09\%), knowledge of the theoretical 
cross-section (0.12\%), experimental systematics (0.15\%) and uncertainty 
in the beam energy (0.04\%). The theoretical cross-section is calculated
using BHLUMI~4.04~\cite{bib:bhlumi}, and a recent assessment of the
theoretical error associated with this program~\cite{bib:bhlumi_err} has 
resulted in a significant decrease in this contribution compared with earlier
analyses. Errors from
the luminosity measurement are included in all the systematic errors
on cross-sections quoted in this paper, and correlations between
measurements arising from the luminosity determination are included
in all fits.

\subsection{Cross-section and Asymmetry Measurements}
\begin{figure}
  \epsfxsize=\textwidth
  \epsfbox[0 0 567 567]{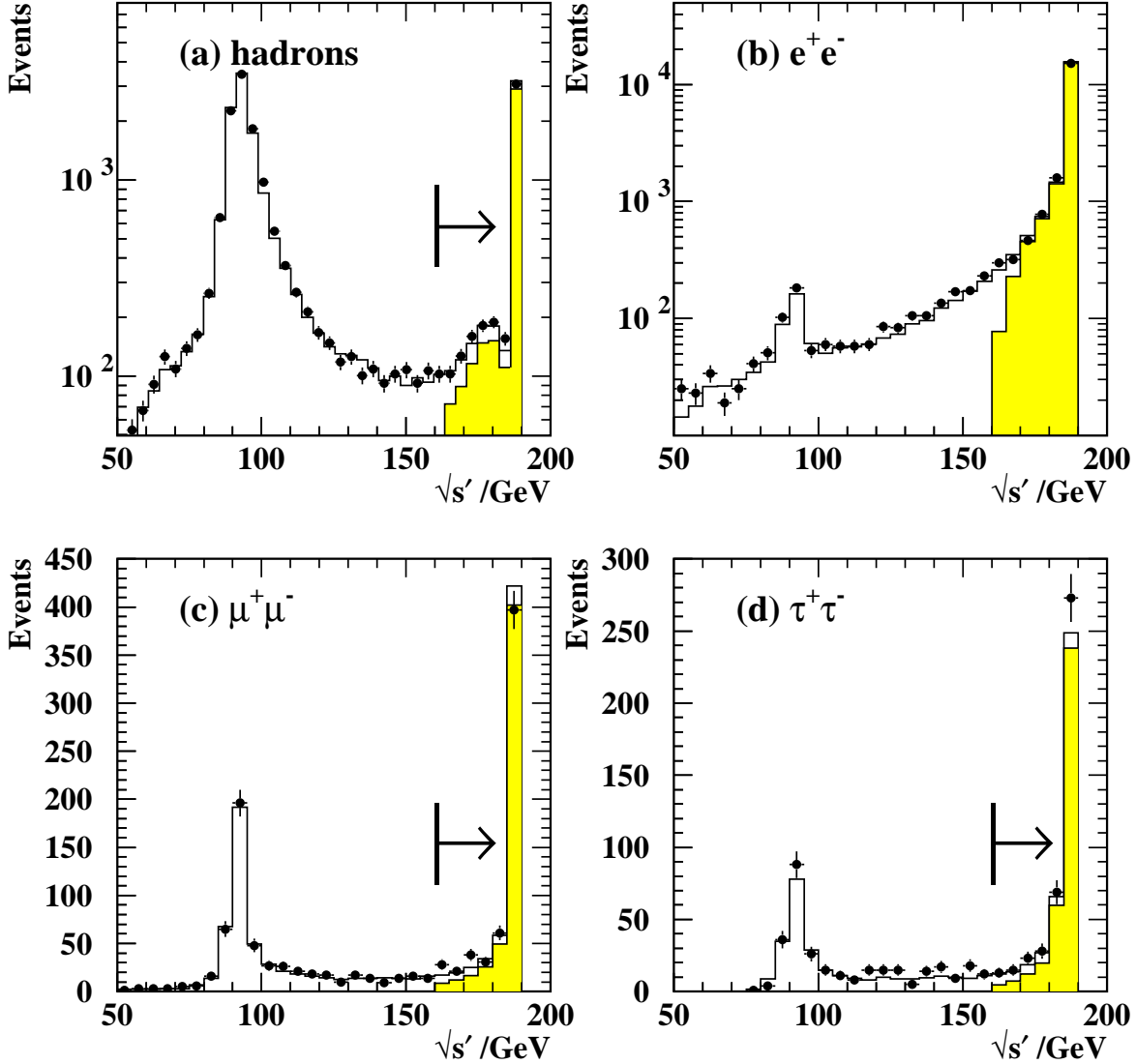}
  \caption
{
 The distributions of reconstructed $\protect\sqrt{s'}$ 
 for (a) hadronic events, (b) electron pair events with $\absctep < 0.9$, 
 $\absctem < 0.9$ and $\thacol < 170\degree$,
 (c) muon pair and (d) tau pair events at 188.63~GeV. In each case, the points 
 show the data and the histogram the Monte Carlo prediction, normalized
 to the integrated luminosity of the data, with the 
 contribution from events with true $s'/s > 0.7225$ shaded in (a), (c) and (d),
 and the contribution from events with $\thacol < 10\degree$ shaded in (b).
 The arrows in (a), (c) and (d) show the position of the cut used to
 select `non-radiative' events. 
}
\label{fig:sp}
\end{figure}

\begin{figure}
  \epsfxsize=\textwidth
  \epsfbox[0 0 567 680]{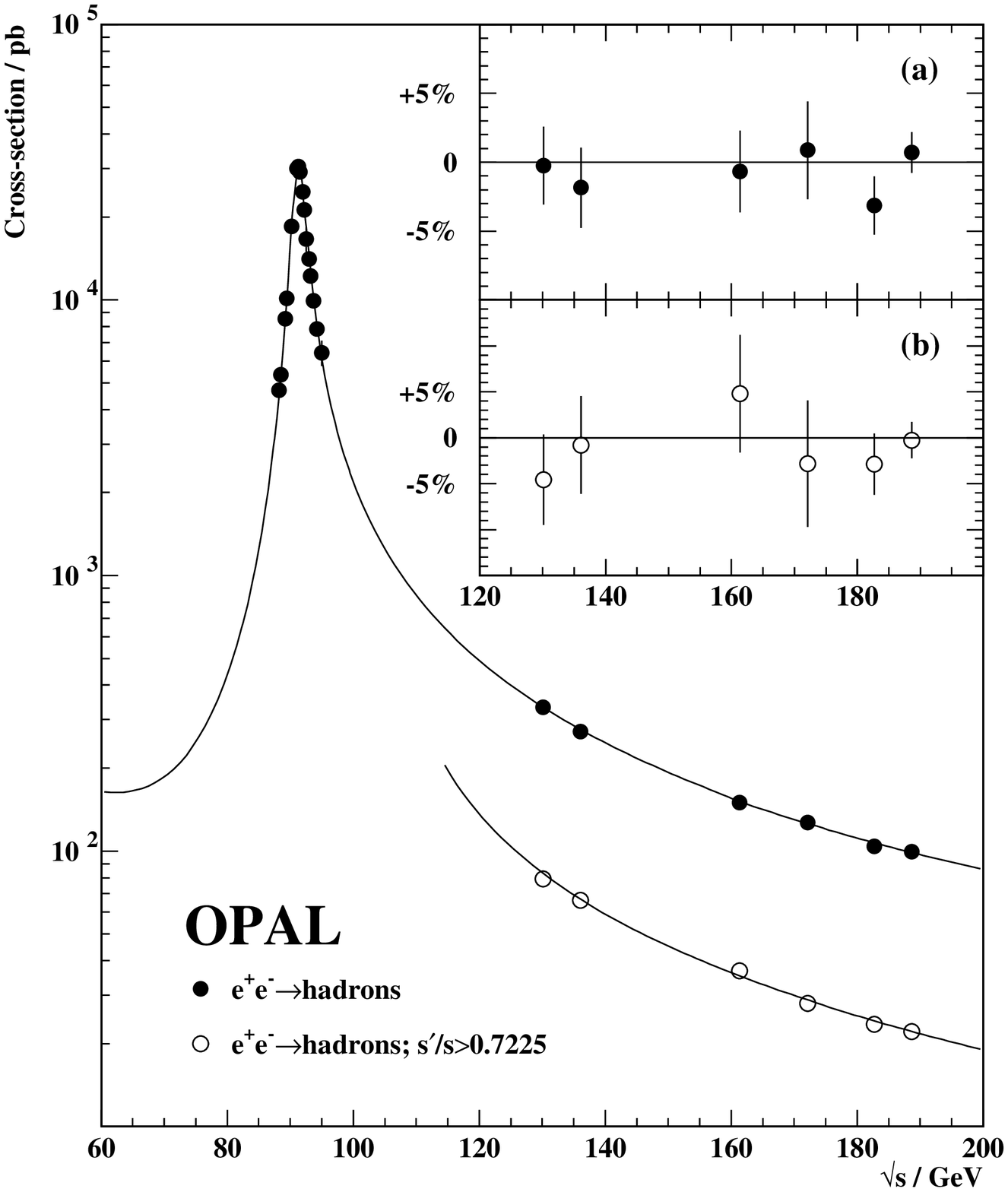}
  \caption
{
    Measured total cross-sections ($s'/s>0.01$) for hadronic events
    at lower energies~\cite{bib:OPAL-LS90,bib:OPAL-LS91,bib:OPAL-LS92,
    bib:OPAL-SM172,bib:OPAL-SM183}, and this analysis. Cross-section
    measurements for $s'/s>0.7225$ from this analysis and 
    from~\cite{bib:OPAL-SM172,bib:OPAL-SM183} are also shown; 
    where necessary, the latter have been corrected from $s'/s > 0.8$ to 
    $s'/s > 0.7225$ by adding the prediction of ZFITTER for this difference 
    before plotting. The curves show the predictions of ZFITTER.
    The insets show the percentage differences between the measured
    values and the ZFITTER predictions for the high energy points
    for (a) $s'/s>0.01$ and (b) $s'/s>0.7225$.
}
\label{fig:mh_xsec}
\end{figure}

\begin{figure}
  \epsfxsize=\textwidth
  \epsfbox[0 0 567 680]{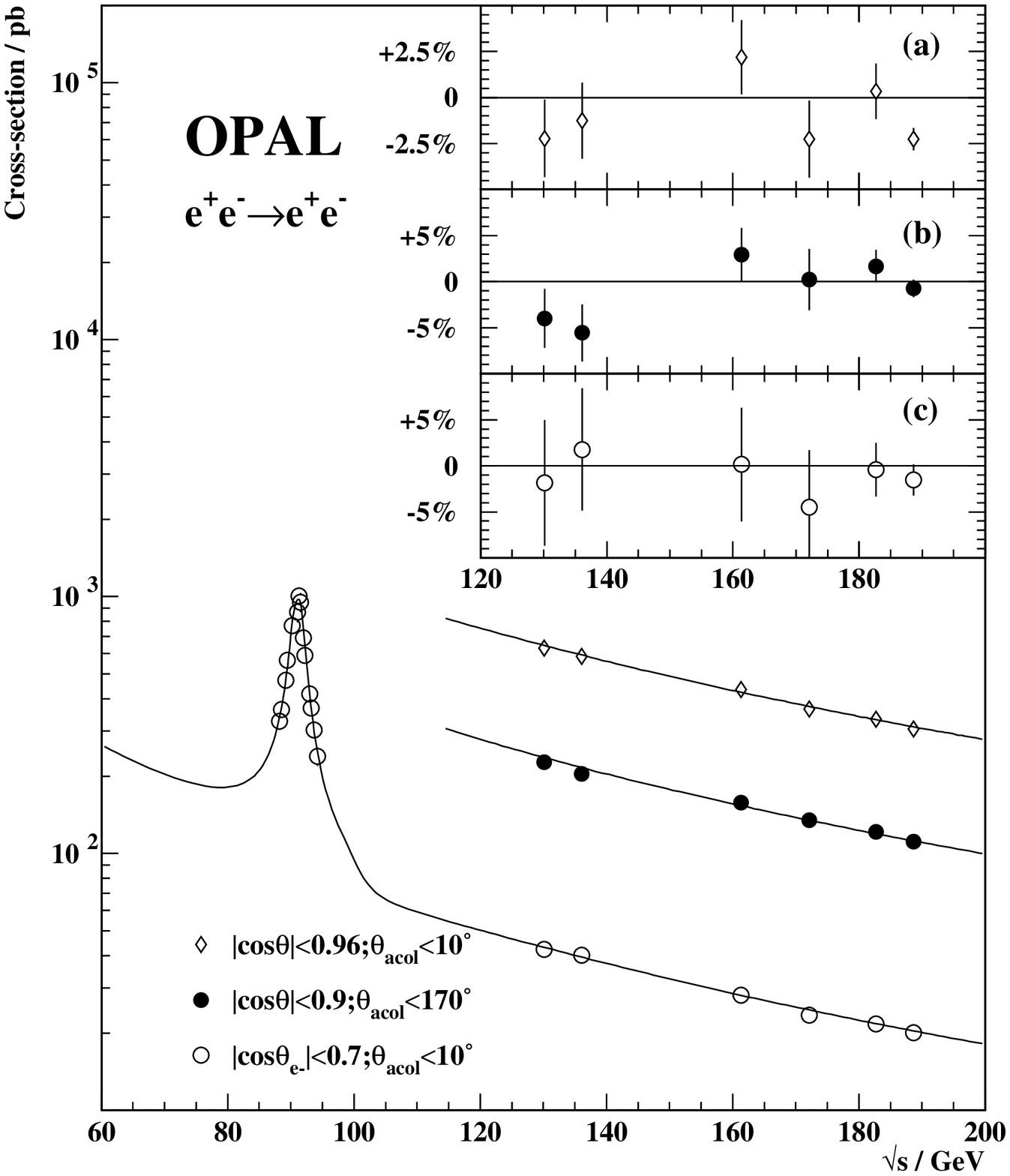}
  \caption
{
    Measured cross-sections for electron pair events
    at lower energies~\cite{bib:OPAL-LS90,bib:OPAL-LS91,bib:OPAL-LS92,
    bib:OPAL-SM172,bib:OPAL-SM183}, and this analysis. 
    The curves show the predictions of ALIBABA.
    The insets show the percentage differences between the measured
    values and the ALIBABA predictions for the high energy points
    for 
    (a) $\absct < 0.96$, $\thacol < 10\degree$,
    (b) $\absct < 0.9$, $\thacol < 170\degree$ and
    (c) $\absctem < 0.7$, $\thacol < 10\degree$.
}
\label{fig:ee_xsec}
\end{figure}

\begin{figure}
  \epsfxsize=\textwidth
  \epsfbox[0 0 567 680]{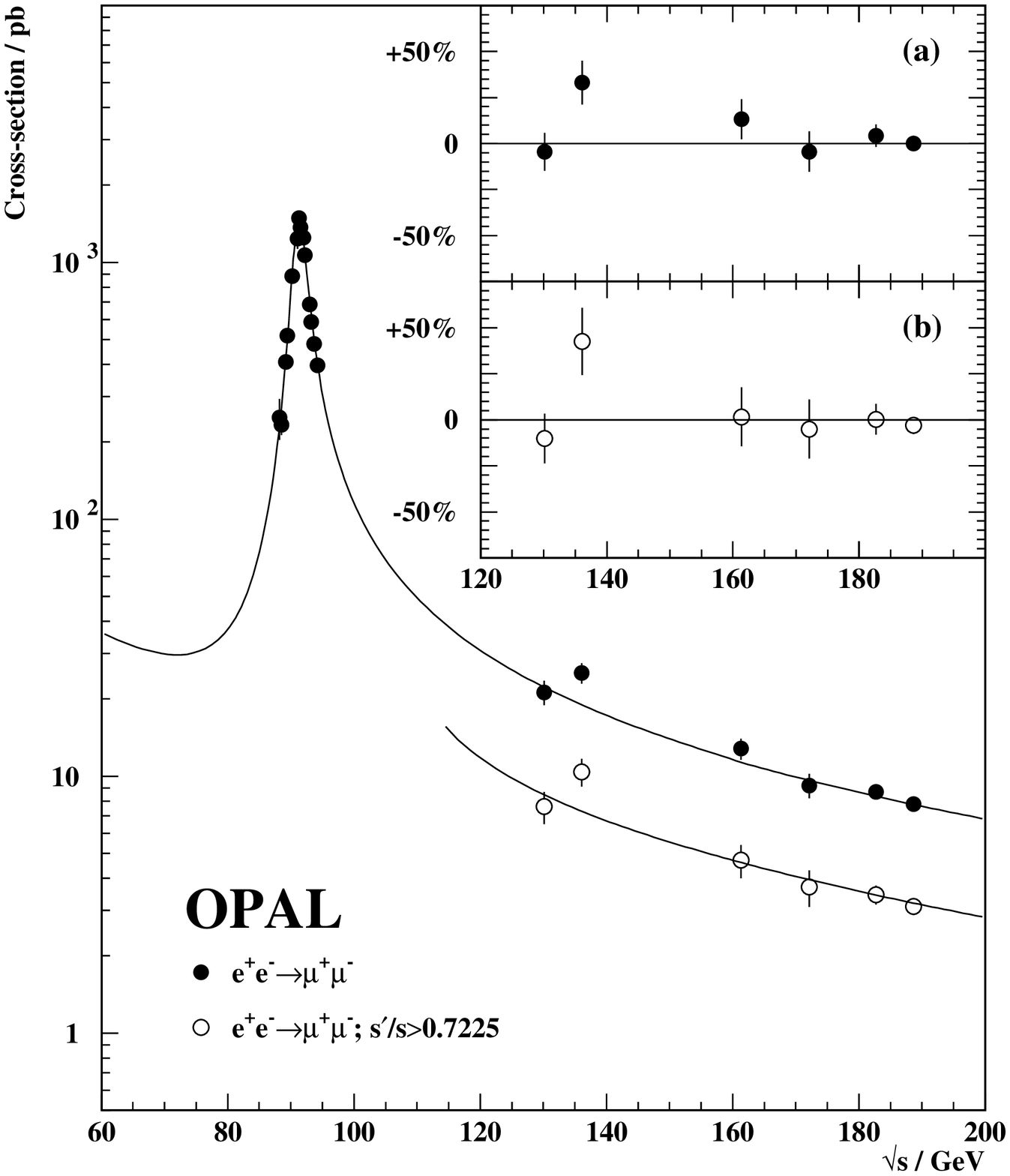}
  \caption
{
    Measured total cross-sections ($s'/s>0.01$) for muon pair events
    at lower energies~\cite{bib:OPAL-LS90,bib:OPAL-LS91,bib:OPAL-LS92,
    bib:OPAL-SM172,bib:OPAL-SM183}, and this analysis. Cross-section
    measurements for $s'/s>0.7225$ from this analysis and 
    from~\cite{bib:OPAL-SM172,bib:OPAL-SM183} are also shown; 
    where necessary, the latter have been corrected from $s'/s > 0.8$ to 
    $s'/s > 0.7225$ by adding the prediction of ZFITTER for this difference 
    before plotting. The curves show the predictions of ZFITTER.
    The insets show the percentage differences between the measured
    values and the ZFITTER predictions for the high energy points
    for (a) $s'/s>0.01$ and (b) $s'/s>0.7225$.
}
\label{fig:mu_xsec}
\end{figure}

\begin{figure}
  \epsfxsize=\textwidth
  \epsfbox[0 0 567 680]{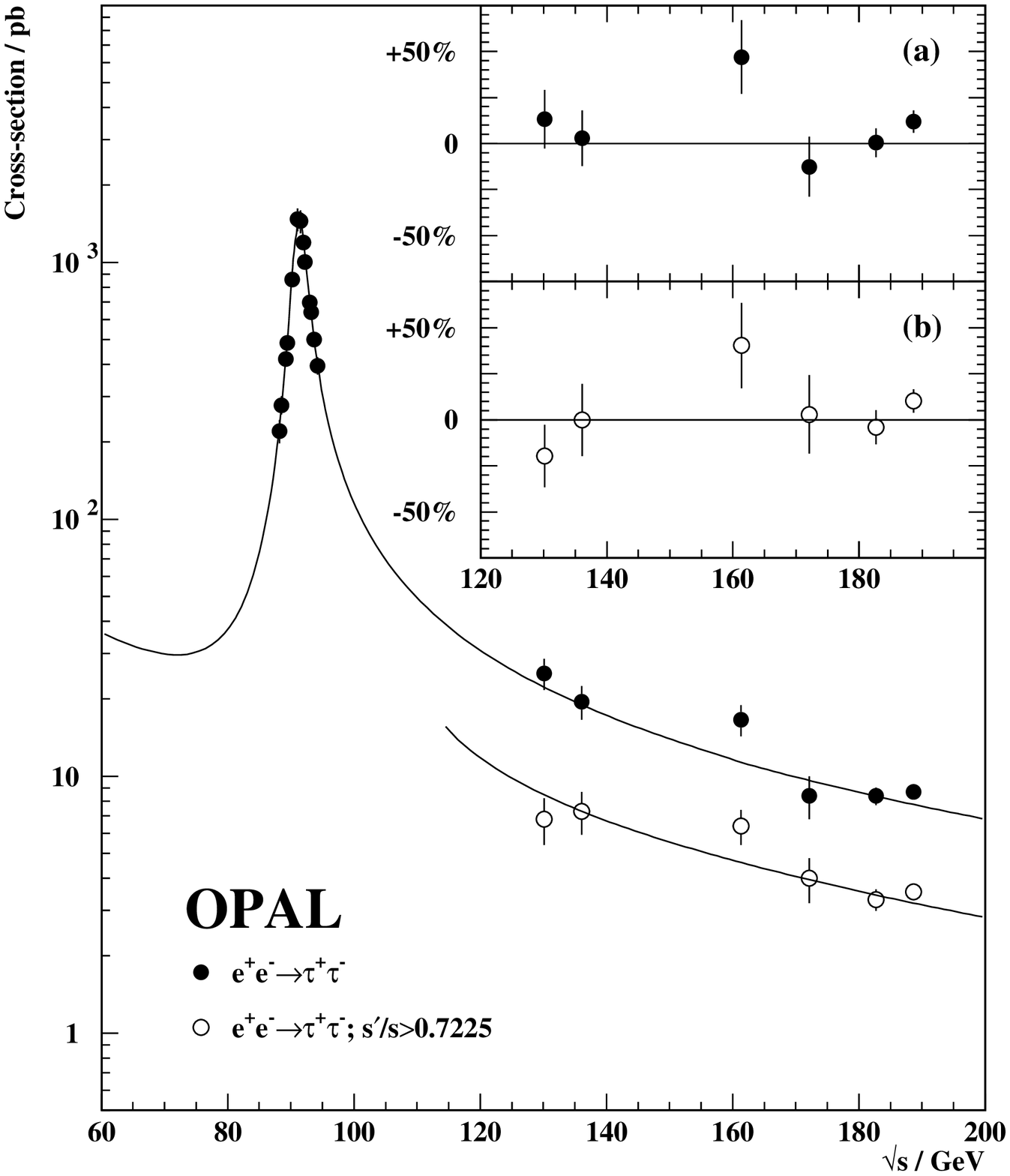}
  \caption
{
    Measured total cross-sections ($s'/s>0.01$) for tau pair events
    at lower energies~\cite{bib:OPAL-LS90,bib:OPAL-LS91,bib:OPAL-LS92,
    bib:OPAL-SM172,bib:OPAL-SM183}, and this analysis. Cross-section
    measurements for $s'/s>0.7225$ from this analysis and 
    from~\cite{bib:OPAL-SM172,bib:OPAL-SM183} are also shown; 
    where necessary, the latter have been corrected from $s'/s > 0.8$ to 
    $s'/s > 0.7225$ by adding the prediction of ZFITTER for this difference 
    before plotting. The curves show the predictions of ZFITTER.
    The insets show the percentage differences between the measured
    values and the ZFITTER predictions for the high energy points
    for (a) $s'/s>0.01$ and (b) $s'/s>0.7225$.
}
\label{fig:tau_xsec}
\end{figure}

Hadronic, \epem, \mumu\ and \tautau\ events were selected using the same 
criteria\footnote{In the selection of muon pairs a minor change was made to 
the cut used to reject cosmic ray events having back-to-back hits in the 
time-of-flight counters. This change reduced the cosmic ray background in the 
selected events, allowing a reduction in the associated systematic error of 
around 40\%.} as at 183~\GeV~\cite{bib:OPAL-SM183}. Distributions of 
$\sqrt{s'}$ for each channel, determined using kinematic fits for hadrons and 
track angles for the lepton pairs as in~\cite{bib:OPAL-SM172}, are shown in 
Fig.~\ref{fig:sp}. Efficiencies, backgrounds and feedthrough of events from 
lower $s'$ into the non-radiative samples were calculated from Monte Carlo 
simulation, and are given in Table~\ref{tab:eff}. Efficiencies determined from
two-fermion Monte Carlo events have been corrected for the effect of the
four-fermion contribution as described in~\cite{bib:OPAL-SM172}. In 
addition, a small correction ($\sim$0.4\%) has been applied to the relevant
electron pair efficiencies to account for tracking problems in regions
of the detector near anode planes of the central jet chamber.
The numbers of selected events and the measured cross-sections are presented 
in Table~\ref{tab:xsec}. The evaluation of the systematic errors is
described in detail below. As well as cross-sections for \qqbar\ events, 
we also present a fully inclusive hadronic cross-section 
$\sigma(\qqbar\mathrm{X})$. This uses the same event selection as is
used for \qqbar\ events but W-pairs are not rejected. The cross-section
therefore includes W-pair (and Z-pair) production with at least one W (Z)
decaying hadronically, but does not include other four-fermion hadronic
events (for example from two-photon processes) which are treated as
background. The energy dependence of the measured cross-section
for each channel is shown in Figs.~\ref{fig:mh_xsec}--\ref{fig:tau_xsec}.

\begin{figure}
\epsfxsize=\textwidth
\epsfbox[0 0 567 680]{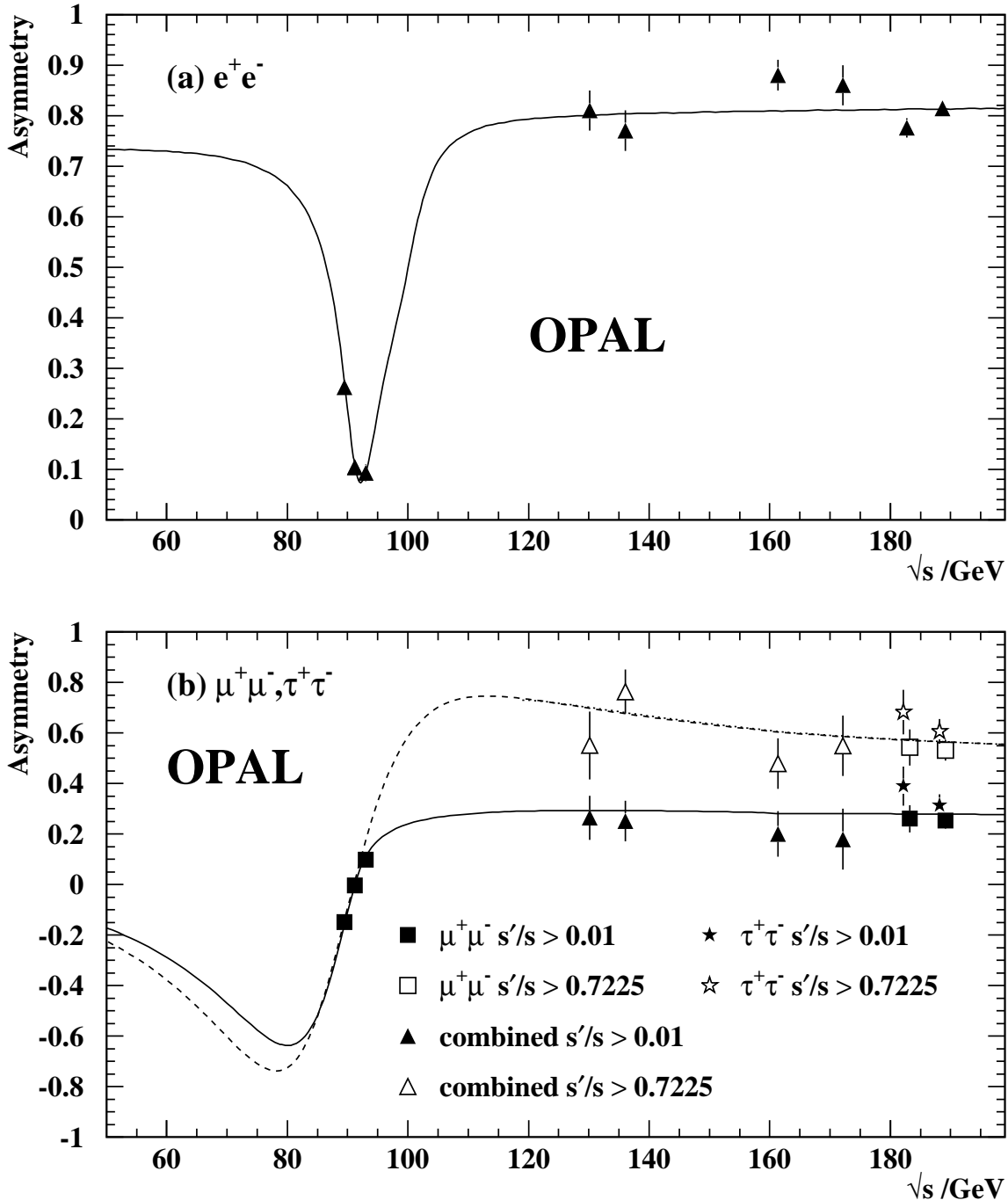}
\caption
{ (a) Measured forward-backward asymmetry for electron pairs with 
  $|\cos\theta_{\Pem}|<0.7$ and $\thacol<10\degree$, as a function
  of $\protect\roots$. The curve shows the prediction of ALIBABA.
  (b) Measured asymmetries for all ($s'/s>0.01$) and non-radiative
  ($s'/s>0.7225$) samples as functions of $\protect\roots$
  for \Pgmp\Pgmm\ and \Pgtp\Pgtm\ events. Some points are plotted
  at slightly displaced values of $\protect\roots$ for clarity.
  The curves show ZFITTER predictions for $s'/s>0.01$ (solid) and
  $s'/s>0.7225$ (dotted), as well as the Born-level expectation
  without QED radiative effects (dashed). The expectation for $s'/s>0.7225$ 
  lies very close to the Born curve, such that it appears 
  indistinguishable on this plot. 
}
\label{fig:afb}
\end{figure}

\begin{figure}
\epsfxsize=\textwidth
\epsfbox[0 0 567 567]{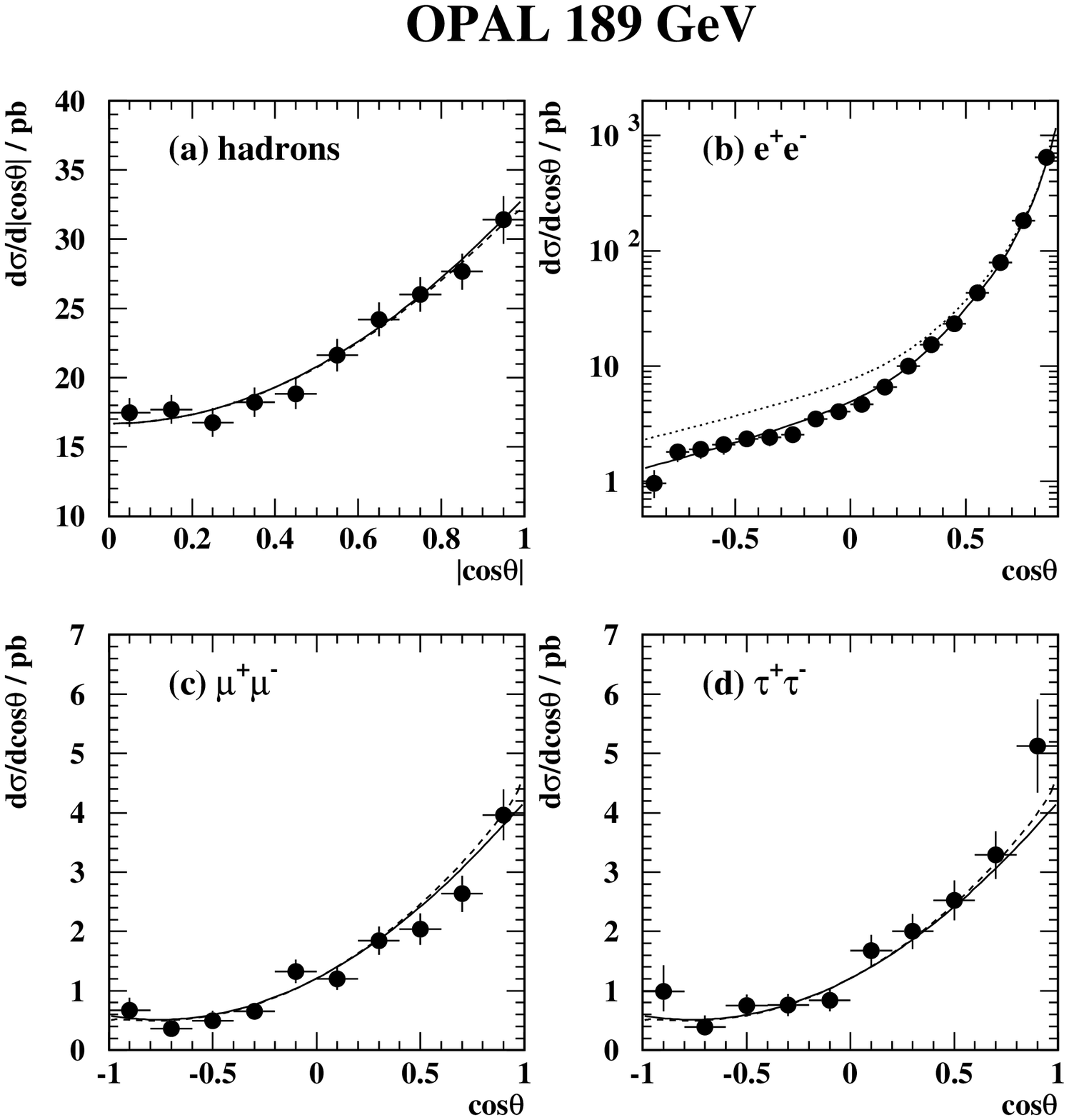}
\caption
{ Angular distributions for (a) hadronic events with $s'/s > 0.7225$,
 (b) \epem\ events with $\thacol < 10\degree$, (c) \mumu\ events with
 $s'/s > 0.7225$ and (d) \tautau\ events with $s'/s > 0.7225$.
 The points show the 189~GeV data, corrected to no interference between
 initial- and final-state radiation in (a), (c) and (d). The solid curve
 in (b) shows the prediction of ALIBABA, while the dotted curve shows
 the prediction with no contribution from $t$-channel Z exchange.
 The curves in (a),(c) and (d) show the predictions of ZFITTER with no 
 interference between initial- and final-state radiation (solid) and with 
 interference (dashed).
}
\label{fig:angdis}
\end{figure}

Measurements of the forward-backward asymmetry for lepton pairs are
given in Table~\ref{tab:afb} and compared with lower energy 
measurements in Fig.~\ref{fig:afb}. 
The values for muon and tau pairs are obtained by averaging the
results measured using the negative particle with those obtained using
the positive particle to reduce systematic effects.
Muon and tau asymmetries are corrected 
to the full angular range by applying a multiplicative correction obtained 
from ZFITTER to the asymmetry measured within the acceptance of the selection
cuts ($\absct < 0.95$ for muon pairs, $\absct < 0.9$ for tau pairs). 
The angular distribution of the primary quark in non-radiative 
hadronic events is given in Table~\ref{tab:mh_angdis}, and the corrected 
angular distributions for the lepton pairs are given in 
Tables~\ref{tab:ee_angdis} and~\ref{tab:mu_angdis}. The angular distributions 
are plotted in Fig.~\ref{fig:angdis}.

All cross-sections and asymmetries except those for \epem\ have been 
corrected for the 
contribution of interference between initial- and final-state radiation as 
described in~\cite{bib:OPAL-SM172}\footnote{The corrections in our
earlier publications~\cite{bib:OPAL-SM172,bib:OPAL-SM183} were based on
the interference cross-sections predicted by ZFITTER version 5.0, which
have subsequently been found by the authors to be a factor of three too big 
for hadronic
final states. We have therefore reduced the interference corrections
applied to the 130--183~GeV hadronic cross-sections by a factor of three
when using these data in the fits described in this paper.}.
The corrections are shown in Table~\ref{tab:ifsr}. 

\subsection{Systematic Studies}
\subsubsection{\bf Hadronic Events}
The selection criteria for hadronic events~\cite{bib:OPAL-SM183,bib:OPAL-SM172}
use the multiplicity of tracks and electromagnetic calorimeter clusters,
the total electromagnetic calorimeter energy and the energy balance along
the beam direction. Events selected as W-pair candidates according to
the criteria of~\cite{bib:OPAL-WW183} are rejected. In selecting
the non-radiative sample a kinematic fit
is used to determine $s'$. The main backgrounds arise from four-fermion
final states.

The systematic errors on the hadronic cross-sections have been
substantially reassessed compared with lower energy analyses. They
are summarized in Table~\ref{tab:mh_syserr}, and the main contributions
are discussed below.

{\bf ISR modelling.} The effect of the modelling of initial-state radiation
on the selection efficiency and $s'$ determination has been estimated by 
comparing the prediction of PYTHIA~\cite{bib:pythia} with that of the 
KK2f~\cite{bib:KK2f} Monte Carlo generator, which has a more complete 
description of initial-state radiation. The difference between the two was
taken as the systematic error.

{\bf Fragmentation modelling.} The effect of the hadronization model on 
the efficiency of the non-radiative selection has been investigated by 
comparing the string fragmentation implemented in PYTHIA with the cluster
model of HERWIG~\cite{bib:herwig}. In order to decouple the effects of
hadronization from differences in initial-state radiation treatment
in the two programs, efficiencies were compared in bins of $s'$.
They were found to agree within the statistical precision of the
test, which was accordingly assigned as the error. In the inclusive
selection, the greatest loss of efficiency comes from the cut on
electromagnetic calorimeter energy. A comparison of jets in Z data
and Monte Carlo showed that this is well simulated by JETSET (but not
by HERWIG). Therefore the systematic error was estimated by changing the energy
scale in the Monte Carlo by the observed difference between data and
JETSET and re-evaluating the efficiency. In addition, the effect of a 
conservative variation of one unit in charged particle multiplicity was 
also taken into account in the inclusive case.

{\bf Detector effects.}
The selection of inclusive events is mainly based on the electromagnetic
calorimeter, and is thus sensitive to the energy scale of the calorimeter,
and any angular dependence of the energy scale. The energy scale in
hadronic events has been studied using data taken at the Z peak in 1998.
The energy scales in data and Monte Carlo agree to better than 0.5\%,
and the systematic error on the efficiency for inclusive
events was estimated by changing the energy scale in data by this amount.
For non-radiative events, a kinematic fit is used to determine $s'$,
which thus depends on jet energies, angles and their errors. Studies of
data taken at the Z peak were again used to assess the uncertainties in
these; in addition, studies of Bhabha events were used to determine
similar uncertainties in the photon energies and angles and their
resolution. The systematic error in the non-radiative events was
determined by changing each of these quantities and re-evaluating $s'$.

{\boldmath \bf $s'$ determination.} Any possible systematic effects in the 
determination
of $s'$ not covered by the above ISR, fragmentation and detector systematics
were assessed by changing the method of calculating $s'$. The algorithm
was changed to allow for only a single radiated photon, the cuts used
to identify isolated photons in the detector were varied, the value of
the resolution parameter used in the jet finding was varied, and for
jets in the forward regions whose energies are poorly measured the kinematic 
fit was compared with the calculation of $s'$ using jet angles. In each case,
the modified algorithm was applied to data and Monte Carlo, and the 
cross-section recomputed. The changes observed were in all cases compatible 
with statistical fluctuations. The largest of these, averaging over data
taken at 189~GeV and at lower energies, was taken as a systematic error.

{\bf WW rejection cuts and WW background.} The systematic error arising
from the effect of the W-pair rejection cuts on the efficiency, and
the uncertainty in the remaining W-pair background, were estimated
in a similar manner to that described in~\cite{bib:OPAL-WW183}.
As a cross-check, we have calculated the hadronic cross-sections
without rejecting W-pair events, by subtracting their expected contribution
instead. The measured values of 99.6$\pm$0.9$\pm$1.2~pb ($s'/s > 0.01$)
and 22.44$\pm$0.42$\pm$0.20~pb ($s'/s > 0.7225$), after correction for 
interference between initial- and final-state radiation, are in good agreement 
with the values in Table~\ref{tab:xsec}.

{\bf Background.} Uncertainties in other background contributions were
estimated by comparing the predictions of various generators. In the
inclusive sample the largest uncertainty arises from the contribution
of two-photon events. At low $Q^2$ the generators PYTHIA and 
PHOJET~\cite{bib:phojet} were compared. At high $Q^2$ the 
TWOGEN~\cite{bib:twogen} program (with the `perimiss' 
option~\cite{bib:OPAL-f2gam}), PYTHIA, HERWIG and PHOJET were used.
In the non-radiative sample the main background arises from 
four-fermion final states. The prediction of grc4f~\cite{bib:grc4f}
was compared with that of EXCALIBUR~\cite{bib:excalibur}.

{\bf Interference.} The uncertainty arising from the removal of the
contribution from interference between initial- and final-state
radiation was estimated as described in~\cite{bib:OPAL-SM172}.

\subsubsection{\bf Electron Pairs}
Electron pair events are required to have low multiplicity
and large energy deposited in the electromagnetic calorimeter. The
`large acceptance' selection ($\absct < 0.96$, $\thacol < 10\degree$) 
does not require tracks associated to electromagnetic calorimeter clusters, 
but all other selections require two of the three highest energy clusters to 
have an associated track. For measurements of the asymmetry and angular 
distribution these tracks are required to have opposite charge.

The systematic errors associated with the electron pair measurements
are summarized in Table~\ref{tab:ee_syserr}. The most important ones
are discussed below.

{\bf Four-fermion contribution.} The full size of the difference in efficiency
from including $s$-channel four-fermion events in the signal definition
was included as a systematic error.

{\bf Multiplicity cuts.} The errors arising from the requirement of
low multiplicity have been estimated by varying the multiplicity cuts used 
by $\pm1$ unit. 

{\bf Calorimeter energy scale and resolution.}  A detailed
comparison between data and Monte Carlo has been made of the energy
scale and resolution of the electromagnetic calorimeter, and the results
of this study used to assess possible effects on the selection efficiency.
Typically the energy scale was varied by 0.3\% and the resolution by
10\% of its value.

{\bf Track requirements.} Matching between tracks and clusters has
been studied using events passing all selection cuts, except that only
one of the three highest energy clusters has an associated track. These
are expected to be mainly $\epem\gamma$ final states where one electron
and the photon lie within the acceptance and $\gamma\gamma$ final states
where one photon has converted in the detector, with small contributions
from other final states. An excess of such events was seen in data 
compared with Monte Carlo expectation; this excess amounted to
approximately 0.8\% of the number of events passing all cuts. Roughly half 
the excess is concentrated in regions of $\phi$ near the anode planes of the 
central jet chamber, and arises from track reconstruction problems in this
region. The other half could arise from track reconstruction problems,
or could arise from problems modelling $\epem\gamma$ or $\gamma\gamma$
events. For each acceptance region we take half the difference between
data and Monte Carlo as a correction to the efficiency to account for
the loss of tracks near jet chamber anode planes, and assign the other
half as a systematic error associated with track requirements.

{\bf Acceptance correction.} Because of the steepness of the angular 
distribution, uncertainties in the determination of $\theta$ are an 
important systematic error. These have been assessed by comparing 
measurements of $\theta$
in the electromagnetic calorimeter with those in the central tracking
chambers and the muon chambers. These studies indicate a possible bias
in the $\theta$ reconstruction of electromagnetic clusters of 
around 1~mrad in the endcap region of the detector. The effect of
the observed biases on the acceptance was calculated using Monte Carlo
events, and assigned as a systematic error associated with the
acceptance correction.

{\bf Background}
The dominant background in the selections including tracks is from
\tautau\ events if a tight acollinearity cut is applied. With a loose
acollinearity cut, $\epem\gamma$ and \epem\epem\ events are also significant.
The systematic error arising from uncertainty in the background has
been assessed by comparing the numbers of events in data and Monte
Carlo which pass all cuts except the cut on total calorimeter energy;
these events are predominantly background. In each acceptance region, 
the numbers agree to within one standard deviation, and the statistical
precision of the test was taken as the associated systematic error.
For the selection which does not use tracks, the only important background
is from $\gamma\gamma$ final states; here we used the statistical
precision of the OPAL cross-section measurement~\cite{bib:OPAL-gg189}
to estimate the systematic error from this background.

The systematic error on the overall normalization of the angular distribution
has been assessed in a similar manner to those on the cross-sections, but
includes an extra contribution from an observed difference of 0.5\%
between data and Monte Carlo in the probability of the two tracks
having opposite charge. The overall error amounts to 0.76\%.

Systematic errors on the asymmetry measurement arise from the effects
of $\theta$ mismeasurement, charge misassigment and background and
efficiency corrections, and amount to 0.005. Even with the current 
statistics this is only half of the statistical error.

\subsubsection{\bf Muon and Tau Pairs}
Muon pair events are required to have two tracks identified as muons.
Background from cosmic ray events is removed using time-of-flight
counters and vertex cuts, and two-photon events are rejected by
placing a cut on the total visible energy.

The selection of tau pair events uses information from the central
tracking detectors and electromagnetic calorimeter to identify events
with two collimated, low multiplicity jets. Background from Bhabha
events is rejected using cuts on the total visible energy and
the electromagnetic calorimeter energy associated with each tau cone.
Two-photon events are rejected using cuts on total visible energy
and missing momentum.

Systematic errors on the muon pair and tau pair cross-sections are 
summarized in Tables~\ref{tab:mu_syserr} and~\ref{tab:tau_syserr}
respectively. They were estimated using similar methods, and the main 
contributions are discussed below.

{\bf Efficiency.} The systematic errors on the efficiencies were evaluated
using high statistics LEP1 data and Monte Carlo samples. The muon pair
or tau pair selection cuts were applied to these samples and the difference 
between 
the number of data events selected and the number expected from Monte Carlo 
was used to estimate the systematic error associated with the efficiency.
For this comparison, it was necessary to relax some of the cuts which 
scale with centre-of-mass energy slightly, so that the efficiency for
events on the Z peak remained high. 

{\bf Cosmic background.} The error due to any remaining cosmic
background in the muon pairs was estimated by varying the time-of-flight or 
vertex cuts by amounts determined from the resolution in the respective 
variables.

{\bf Other backgrounds}. The main backgrounds in the muon pairs
arise from \epem\mumu, \tautau\ and leptonic four-fermion
final states. The largest background in the tau pairs arises from Bhabha 
events. Other important backgrounds arise from \epem\epem\ and \epem\tautau\ 
final states.
Backgrounds were studied by considering distributions of 
selection variables after all cuts except the one on that variable. The 
numbers of events in data and Monte Carlo were compared for a region enriched 
in a particular background, and the difference, or its statistical error, 
used to estimate the systematic error from that background source. For 
example, the \epem\mumu\ background in the muon pairs was studied using 
the distribution of visible energy; the Bhabha background in the tau
pairs was estimated using distributions of total visible energy with
the cuts on energy relaxed.
For backgrounds which cannot be studied in this way,
we conservatively assume an error of 50\%.

{\bf Interference.} The uncertainty arising from the removal of the
contribution from interference between initial- and final-state
radiation was estimated as described in~\cite{bib:OPAL-SM172}.

\section{Comparison with Standard Model Predictions}    \label{sec:sm}
The cross-section and asymmetry measurements at 189~GeV are compared
with the Standard Model predictions in Tables~\ref{tab:xsec} 
and~\ref{tab:afb} respectively. 
Figures~\ref{fig:mh_xsec}--\ref{fig:tau_xsec} show cross-sections, 
for both inclusive and non-radiative events, as a function of $\roots$, 
while Fig.~\ref{fig:afb} shows the measured asymmetry values.
The Standard Model predictions
are calculated using ALIBABA~\cite{bib:alibaba} for the \epem\ final
state and ZFITTER~\cite{bib:zfitter} for all other final states;
in this paper we use ZFITTER version 6.10 with input parameters as 
in~\cite{bib:OPAL-SM172}, except that the mass of the Higgs boson is
set to 175~GeV, roughly midway between the experimental lower 
bound~\cite{bib:mH_exp} and the 95\% confidence level upper limit from
electroweak fits~\cite{bib:mH_ew}. Values predicted by the 
TOPAZ0~\cite{bib:topaz0}
program are higher than those predicted by ZFITTER by about 0.03\% (0.2\%)
for non-radiative (inclusive) muon and tau pair cross-sections, and
by about 0.6\% (0.8\%) for the non-radiative (inclusive) hadronic
cross-section. A major theoretical uncertainty on these cross-sections 
arises from the contribution of virtual pairs. By comparing the predictions 
of ZFITTER for the sum of real and virtual pairs with that of the 
four-fermion Monte Carlo grc4f for real pairs, we estimate this contribution 
to be around --0.6\% of the muon or tau pair cross-section, independent of
the $s'$ cut. In the fits described below we therefore assign the
full size of this contribution, 0.6\%, as the theoretical error on
non-radiative muon and tau pair cross-sections. For non-radiative hadrons,
we combine the size of the virtual pair contribution with the difference
seen between the ZFITTER and TOPAZ0 programs to give a theoretical
error of 0.8\%. In the case of electron pairs, we have compared the
predictions of ALIBABA with those of TOPAZ0, and also with those of
the BHWIDE~\cite{bib:bhwide} Monte Carlo program. Based on these comparisons, 
we assign a theoretical error of 2\% to electron pairs in the fits
below. For the muon and tau pair asymmetries, we use a theoretical
error of 0.005, based on comparisons between ZFITTER, TOPAZ0 and
the KK2f Monte Carlo program.
The agreement between the measured cross-sections and Standard Model
predictions is generally good.

The angular distributions for all channels at 189~GeV are compared with 
Standard Model predictions in Fig.~\ref{fig:angdis}. In the case
of electron pairs, we also show the distribution which would be expected
if there were no contribution from the $t$-channel Z-exchange diagram.
We clearly see that the contribution of this diagram is necessary
to reproduce the measured distribution.

\begin{figure}
\epsfxsize=\textwidth
\epsfbox[0 0 567 567]{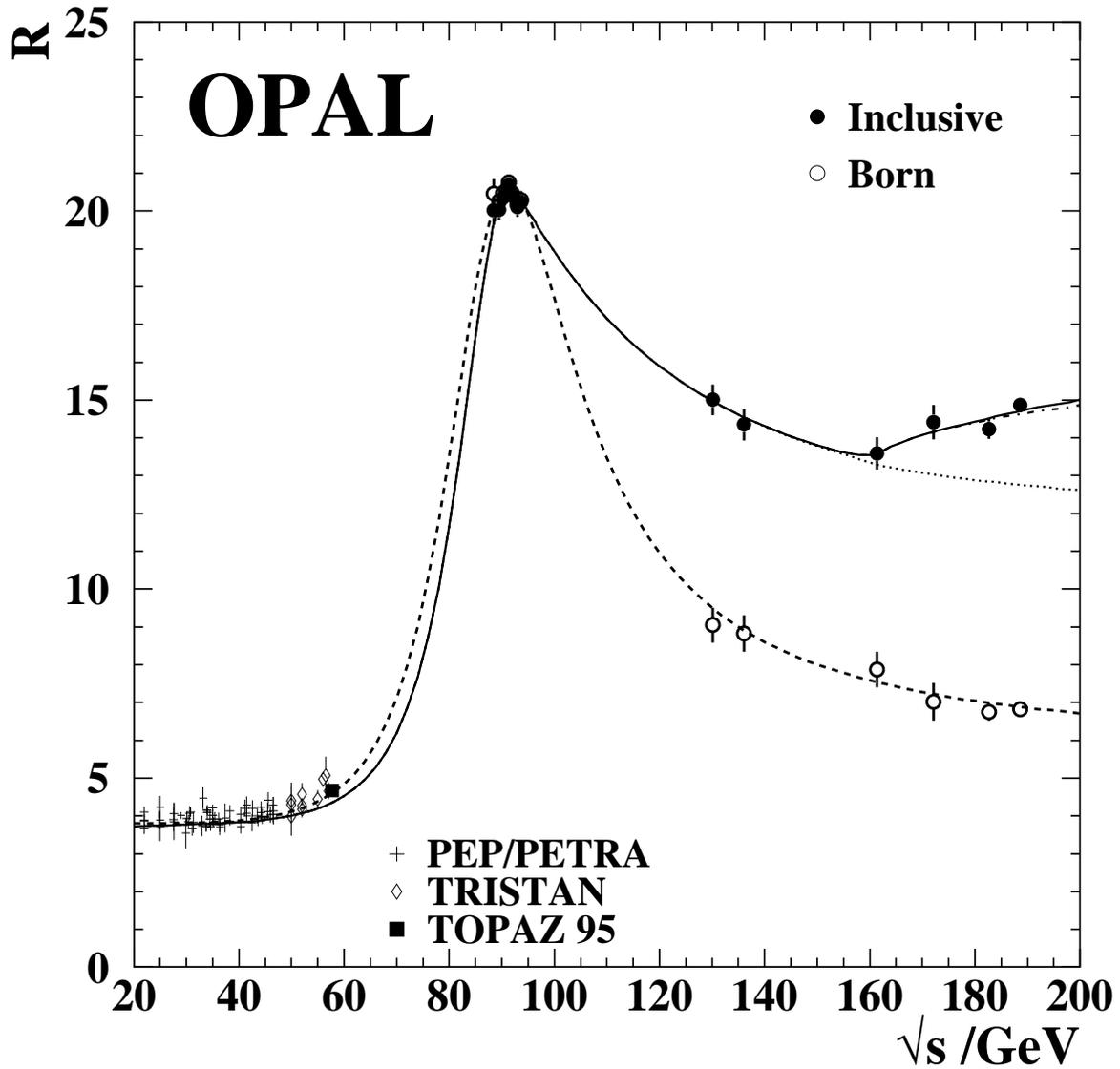}
\caption
{Ratio of measured hadronic cross-sections to theoretical muon
 pair cross-sections as a function of centre-of-mass energy. 
 Values are shown for the inclusive cross-section, $\sigma(\qqbar\mathrm{X})$
 and for the Born level cross-section. The dotted and dashed curves show 
 the predictions of ZFITTER for these cross-sections, 
 while the solid curve also includes the contributions from W-pairs 
 calculated using GENTLE~\cite{bib:gentle} and from Z-pairs calculated
 using FERMISV~\cite{bib:fermisv}. The dot-dashed curve is the total
 excluding the Z-pair contribution. Measurements at lower energies are from 
 references~\cite{bib:OPAL-SM183,bib:OPAL-SM172,bib:OPAL-LS90,bib:OPAL-LS91,
 bib:OPAL-LS92,bib:rdata}.
}
\label{fig:rplot}
\end{figure}

In Fig.~\ref{fig:rplot} we show the ratio of measured hadronic 
cross-sections to theoretical muon pair cross-sections as a function
of centre-of-mass energy for two cases. In the first case
the numerator of this ratio is the inclusive $\qqbar\mathrm{X}$ cross-section,
in the second case it is the non-radiative \qqbar\ cross-section
corrected to the Born level\footnote{Born level means the
cross-section obtained from the improved Born approximation before
convolution with QED radiation; electroweak and QCD corrections
are included.}. In each case the denominator is the
corresponding muon pair cross-section calculated using ZFITTER. The inclusive
ratio clearly shows the effect of \WW\ production.

\subsection{\boldmath Energy Dependence of \alphaem} \label{sec:alphaem}
\begin{figure}
\epsfxsize=\textwidth
\epsfbox[0 0 567 567]{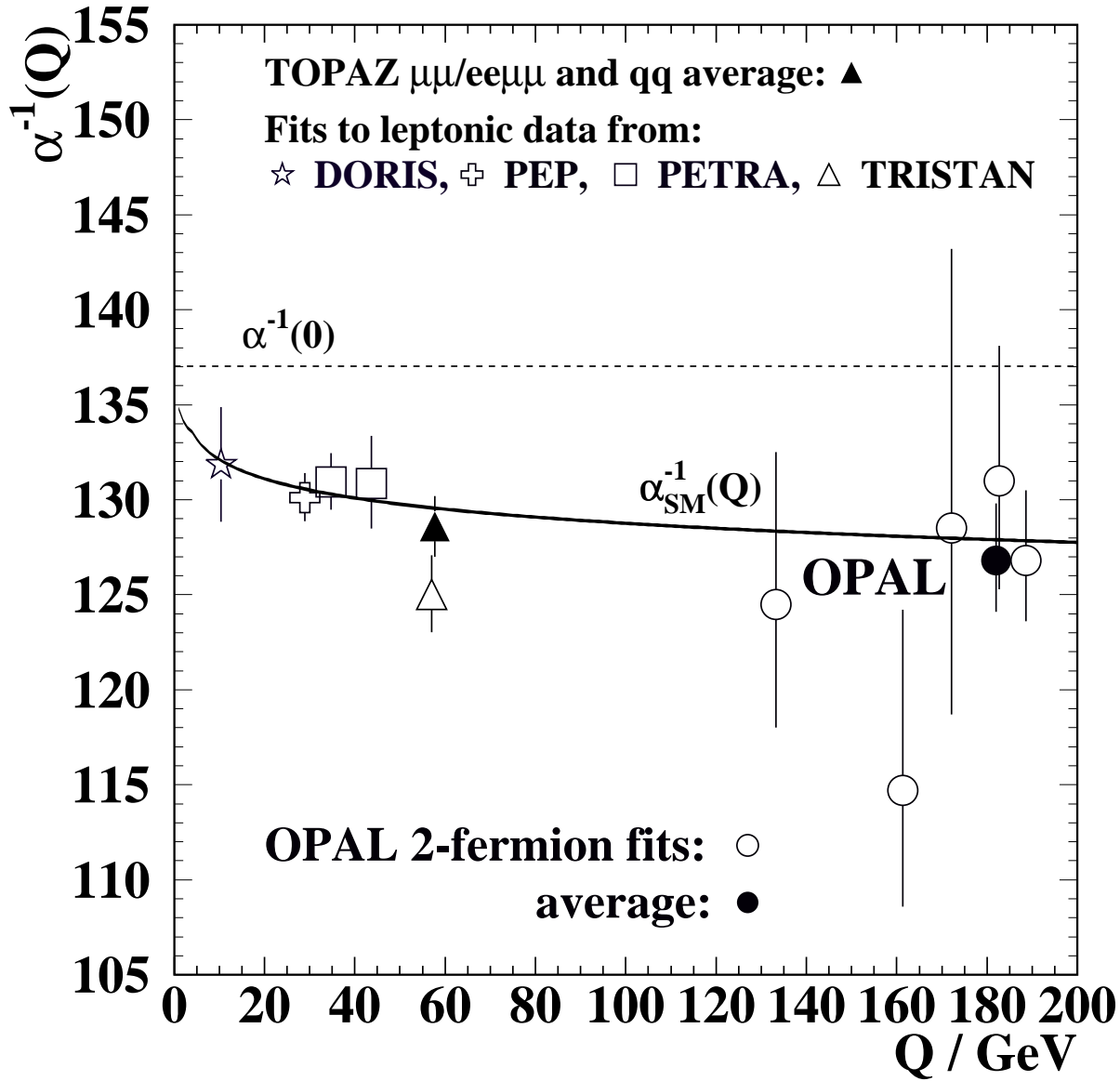}
\caption
{Fitted values of $1/\alphaem$ as a function of $Q$, which is
 $\protect \sqrt{s}$ for the OPAL fits. The open circles show the
 results of fits to OPAL data at each centre-of-mass energy, the closed
 circle the result of the combined fit in which \alphaem\ runs with a
 slope corresponding to its fitted value. The OPAL results at 130--183~GeV
 are from~\cite{bib:OPAL-SM172,bib:OPAL-SM183}. Values obtained by the TOPAZ 
 experiment~\cite{bib:alrun} and from fits to measurements of leptonic
 cross-sections and asymmetries at the DORIS, PEP, PETRA and TRISTAN
 \epem\ storage rings~\cite{bib:MK_alphaem} are also shown. All
 measurements rely on assuming the Standard Model running of \alphaem\
 up to the $Q^2$ of the luminosity measurements, 
 $Q_{\mathrm{lumi}}\sim 5$~GeV. 
 The solid line shows the Standard 
 Model expectation, with the thickness representing the uncertainty, 
 while the value of 1/$\alphaem(0)$ is shown by the dashed line. 
}
\label{fig:alphaem}
\end{figure}

Non-radiative cross-section and asymmetry measurements have been used to
measure the electromagnetic coupling constant \alphaem\ at LEP2 energies,
as described in~\cite{bib:OPAL-SM172,bib:OPAL-SM183}. We form
the $\chi^2$ between measured values and the Standard Model 
predictions calculated as a function of $\alphaem(\sqrt{s})$ using
ZFITTER, with all other ZFITTER input parameters fixed. Correlations
between measurements are fully taken into account. We perform two fits.
The first one uses only the measurements of hadronic, \mumu\ and \tautau\
cross-sections and the combined muon and tau asymmetry values, for
$s'/s > 0.7225$, presented here. The second fit also includes data
at 130--183~GeV~\cite{bib:OPAL-SM183,bib:OPAL-SM172}; in this combined
fit $\alphaem$ runs with energy with a slope corresponding to the 
fitted value. The results of both fits are given in Table~\ref{tab:alphaem1},
and measured values of $\alphaem$ are shown in Fig.~\ref{fig:alphaem}.
They are consistent with the Standard Model expectation. The value
of $1/\alphaem$ obtained from the combined measurements is 3.4 standard 
deviations below the low energy limit of 
137.0359979$\pm$0.0000032~\cite{bib:CAGE}.

The combined fit described above uses measurements of cross-sections which 
depend on the measurement of luminosity, which itself assumes the
Standard Model running of $\alphaem$ from $(Q^2 = 0)$ to typically
$Q^2 = (3.5~\GeV)^2$, where $1/\alphaem\simeq 134$. Therefore it
measures the running of $\alphaem$ only from $Q_{\rm lumi}\simeq 3.5~\GeV$ 
onwards. As before, to become independent of the luminosity measurement,
we have repeated the combined fit replacing the 
cross-sections for hadrons, muon and tau pairs 
with the ratios $\sigma(\mu\mu)/\sigma(\qqbar)$ and 
$\sigma(\tau\tau)/\sigma(\qqbar)$. This is possible since, above
the Z peak, hadrons and leptons have very different sensitivity to
\alphaem\ as discussed in~\cite{bib:OPAL-SM172}.
The result of this fit is
$1/\alphaem(181.94~\mathrm{GeV}) = 126.2^{+3.5}_{-3.2}$, with a
$\chi^2$ of 11.8 for 18 degrees of freedom. The value is close to
that obtained from the cross-section fit but with somewhat larger errors. 
The difference in $\chi^2$ between the best fit and the assumption
that $\alphaem$ does not run with energy but is fixed at the low
energy limit is 7.88. If $\alphaem$ did not run with energy,
the probability of measuring $1/\alphaem$ = 126.2 or lower 
would be 0.25\%, thus demonstrating the running of $\alphaem$ from
$(Q^2 = 0)$ to LEP2 energies. 
This measurement of $\alphaem$ is independent of low-mass hadronic loops 
and nearly independent of the mass of the Higgs boson and $\alphas$; 
it can be scaled to the mass of the Z, giving 
$1/\alphaem(91.19~\mathrm{GeV}) = 127.4^{+3.2}_{-2.9}$.

\section{Constraints on New Physics}     \label{sec:new_phys}
Deviations of the measured data from Standard Model predictions would be
an indication of new physics processes. The good agreement between data and 
the Standard Model places severe constraints on the energy scale of new
phenomena. In this section we report the results of three analyses
in which limits are set on various new physics processes. Firstly we
consider a four-fermion contact interaction. This offers an appropriate
framework for searching for the effects of the exchange of a new particle 
with mass $\mX\gg\sqrt{s}$. Limits on the energy scale $\Lambda$ are 
presented for various models. For lower mass ranges, \mbox{$\sqrt{s}$
\raisebox{2pt}{\mbox{$<$}}\makebox[-8pt]{\raisebox{-3pt}{$\sim$}\,} $\
\, \mX < \Lambda$,} propagator and width effects must be taken into
account. The results of a search for heavy particles which couple
to leptons are reported. Finally we present the results of a search
for the indirect effects of the gravitational interaction in extra
dimensions on the \mumu\ and \tautau\ final states.

\subsection{Limits on Four-fermion Contact Interactions}  \label{sec:ci}
A very general framework in which to search for the effect of new
physics is the four-fermion contact interaction. In this 
framework~\cite{bib:Eichten} the Standard Model Lagrangian for 
$\epem\to\ffbar$ is extended by a term describing a new effective interaction 
with an unknown coupling constant $g$ and an energy scale $\Lambda$:
\begin{eqnarray}\label{eq-contact}
{\cal L}^{\mathrm{contact}} & = 
         & \frac{g^2}{(1 + \delta)\Lambda^2}
         \sum_{i,j=\mathrm{L,R}}\eta_{ij}[\bar{\Pe}_i\gamma^{\mu}{\Pe}_i]
                                       [\bar{\rm f}_j\gamma_{\mu}{\rm f}_j] ,
\end{eqnarray} 
where $\delta = 1$ for $\epem\to\epem$ and $\delta = 0$ otherwise.
Here $\mathrm{e_L} (f_\mathrm{L})$ and $\mathrm{e_R} (f_\mathrm{R})$ 
are chirality projections of electron (fermion) 
spinors, and $\eta_{ij}$ describes the chiral structure of 
the interaction. The parameters $\eta_{ij}$ are free in these models, but
typical values are between $-1$ and $+1$, depending on the type of theory 
assumed~\cite{bib:contacttable}. Here we consider the same set of 
models as in~\cite{bib:OPAL-SM172}.

We have repeated the analysis described in~\cite{bib:OPAL-SM172}, including
the measurements of the angular distributions for the non-radiative
\mbox{$\eetoee$,} $\eetomumu$, $\eetotautau$ processes and the
non-radiative cross-section for $\eetoqq$
at 189~GeV presented here. As before, we used a maximum
likelihood fit in the case of the lepton angular distributions, and
a $\chi^2$ fit for the hadronic cross-sections. Radiative
corrections to the lowest order cross-section were taken into account
as described in~\cite{bib:OPAL-SM172}. Theoretical uncertainties in
the Standard Model predictions were taken into account as discussed
in Section~\ref{sec:sm}. Limits on the energy scale $\Lambda$ were extracted 
assuming $g^2/4 \pi = 1$.

\begin{figure}
\begin{center}
\epsfxsize=0.86\textwidth
\epsfbox{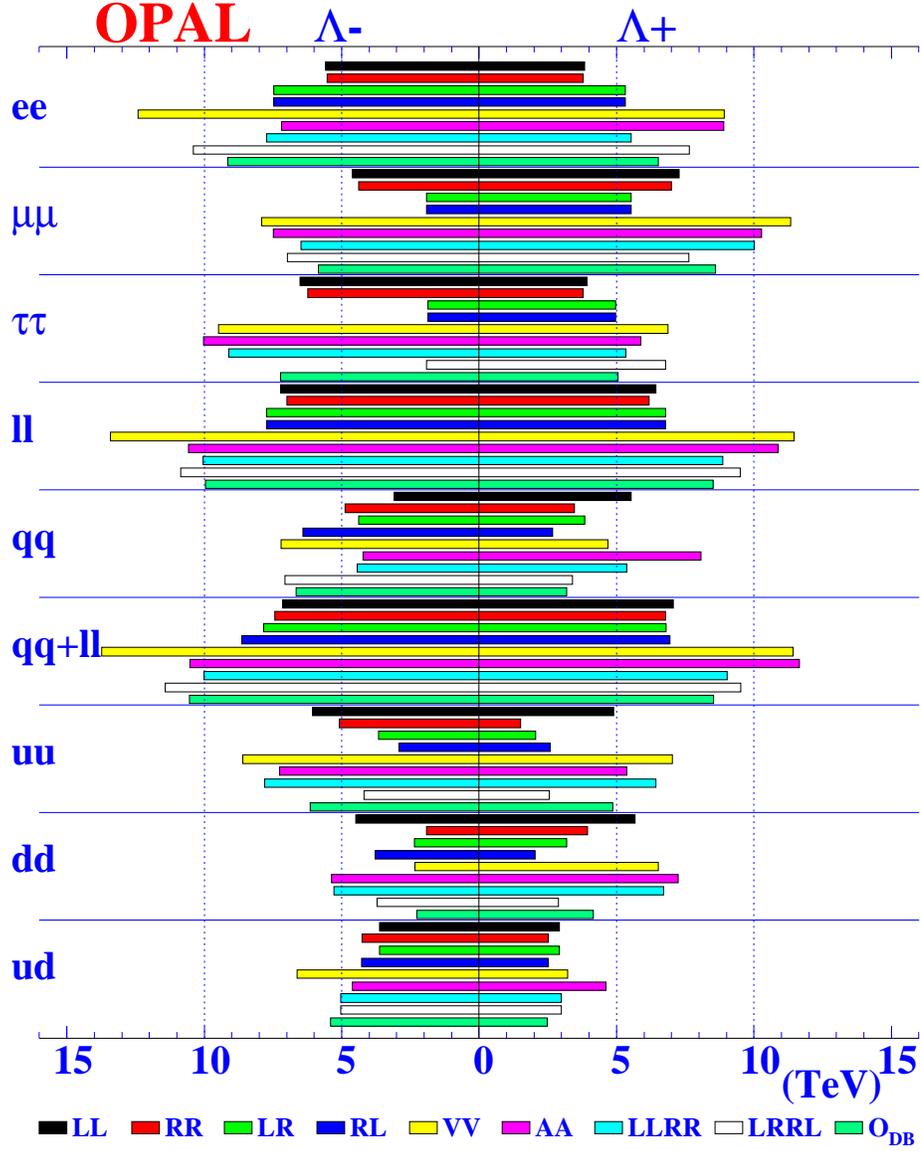}
\caption{95\% confidence level limits on the energy scale $\Lambda$
resulting from the contact interaction fits. For each channel, the
bars from top to bottom indicate the results for models LL to 
$\overline{\cal{O}}_{\mathrm{DB}}$ in the order given in the key.
}
\label{fig:ccres} 
\end{center}
\end{figure}

The results are shown in Table~\ref{tab:ccres} and illustrated
graphically in Fig.~\ref{fig:ccres}; the notation for
the different models is identical to~\cite{bib:OPAL-CI,bib:OPAL-SM172}.
The two sets of values $\Lambda_+$ and $\Lambda_-$ shown in 
Table~\ref{tab:ccres} correspond to positive and negative values of 
$\varepsilon = 1/\Lambda^2$ respectively, reflecting the two possible signs 
of $\eta_{ij}$ in Equation~\eqref{eq-contact}.
As before, the data are particularly sensitive to the 
VV and AA models; the combined data give limits on $\Lambda$ in the
range 10--14~TeV for these models. For the other models the limits
generally lie in the range 7--10~TeV. These limits are roughly
1--2~TeV above those from the 130--183~GeV data alone. 

Contact interactions involving quarks have also been studied in ep and
pp collisions, where limits comparable to our values are
found~\cite{bib:CI-H1,bib:CI-CDF}. Atomic physics parity violation
experiments can place higher limits ($\simeq$ 15~TeV~\cite{bib:CI-atomic})
on models of eeuu and eedd contact interactions which violate parity.

\subsection{Limits on Heavy Particles Coupling to Leptons}    \label{sec:lq}
In this section we present the results of a search under
the explicit assumption that any new phenomena are due to a heavy
particle which couples to leptons. Although we use specific
particles in the analysis presented below, the results are generally 
applicable for any heavy particle with similar properties.

Examples of particles which couple to leptons are sneutrinos with
$R$-parity violating couplings. These couplings are given by the term 
$\lambda_{ijk}L^i_{\rm L}L^j_{\rm L}\overline{E}^k_{\rm R}$  
of the superpotential~\cite{bib:rpvsup}, where the indices $i,j,k$
denote the family of the particles involved, $L^i_{\rm L}$ and $L^j_{\rm L}$ 
are the SU(2) doublet lepton superfields and $\overline{E}^k_{\rm R}$ denotes 
an antilepton singlet superfield. The couplings $\lambda_{ijk}$ are 
non-vanishing only for $i < j$, so at least two different generations of 
leptons are coupled in purely leptonic vertices.

Sneutrinos may contribute to leptonic cross-sections via both $s$-channel
and $t$-channel diagrams, depending on the type of sneutrino and the
final state considered. Processes involving an $s$-channel diagram
lead to resonant behaviour when the centre-of-mass energy is near
the sneutrino mass, and hence more stringent limits can be set than
for processes involving only $t$-channel diagrams. Here we consider
two typical cases involving an $s$-channel diagram:
\begin{itemize}
 \item the presence of a $\snu_\tau$ which interacts via the 
       coupling $\lambda_{131}$ giving rise to a change in the \epem\
       cross-section via an $s$-channel and a $t$-channel process;
       the limits obtained for this case could equally apply to a $\snu_\mu$
       interacting via the coupling $\lambda_{121}$;
 \item a $\tilde \nu_\tau$ with the couplings $\lambda_{131}$ and  
       $\lambda_{232}$ both different from zero. In the analysis both 
       couplings are assumed to be of equal size\footnote{Both 
       couplings violate conservation of the same lepton flavours so that this scenario 
       is compatible with the experimental observation of lepton 
       number conservation.}. Such a scenario gives rise to a modified \mumu\
       cross-section due to an $s$-channel exchange of the sneutrino. 
\end{itemize}
To calculate the differential cross-sections for these processes we use the 
formulae in~\cite{bib:rpvsnu}, taking radiative corrections into account as
in the contact interaction analysis.

In each case we use a maximum likelihood fit of the model prediction to data
and extract 95\% confidence level limits on the coupling
as a function of the sneutrino mass. We include in the fits the data at 
189~GeV presented here and the data at 130--183~GeV presented 
in~\cite{bib:OPAL-SM172,bib:OPAL-SM183}. We use the full
$\sqrt{s'}$ distributions as described in~\cite{bib:OPAL-SM183} in order
to improve the sensitivity at values of sneutrino mass between the 
centre-of-mass energies of LEP. The inclusion of asymmetry values has a very 
small effect on the limits, therefore we present only limits excluding 
asymmetry measurements in order not to lose generality. 

\begin{figure}
\begin{center}
\epsfxsize=\textwidth
\epsfbox[0 150 595 742]{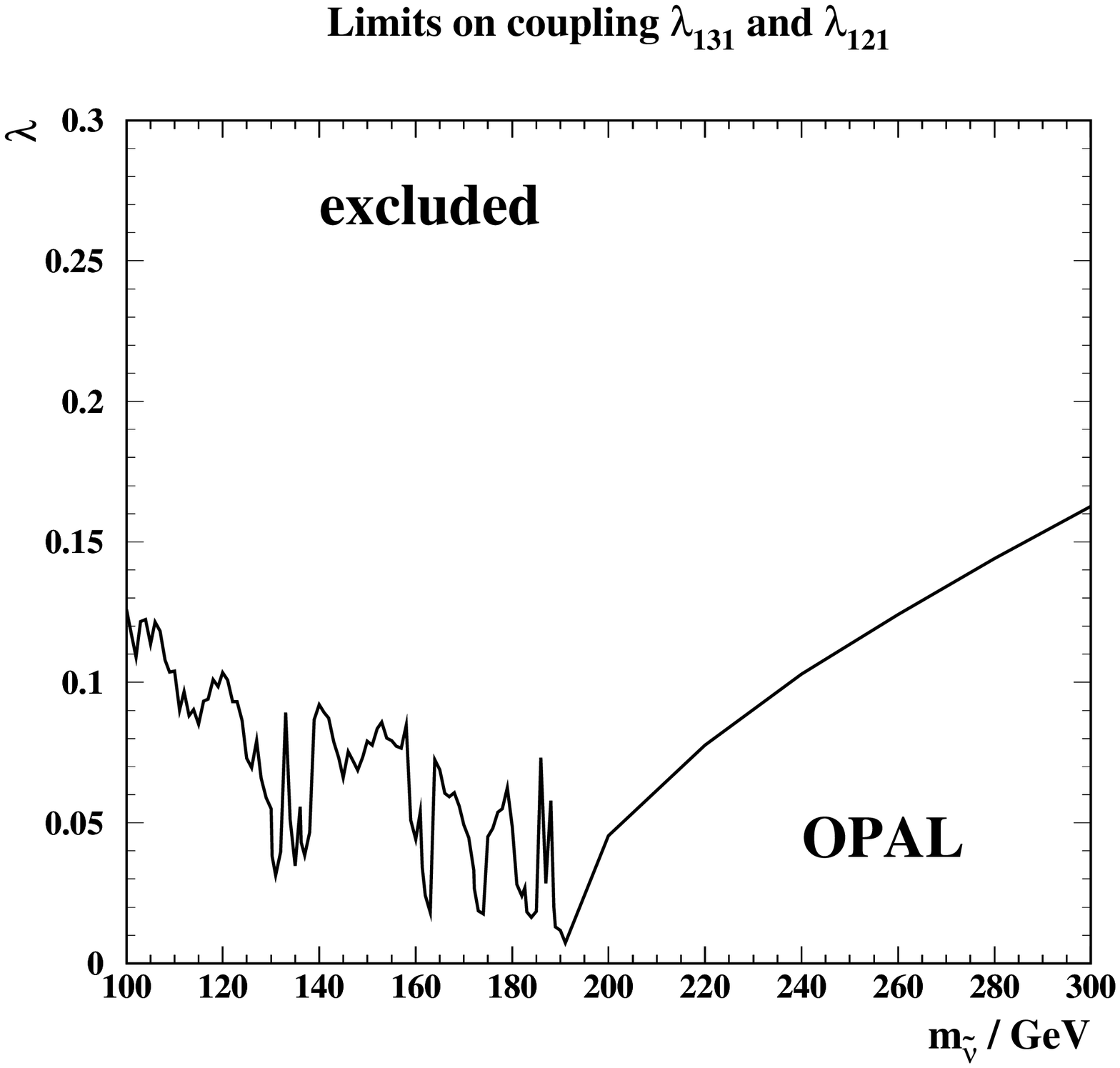}
\caption{95\% confidence exclusion limits on $\protect\lambda_{131}$ 
 (or $\protect\lambda_{121}$) as a function of sneutrino mass 
 $\protect m_{\snu}$, derived from \epem\ $s'$ distributions. The region 
 above the solid line is excluded.
 }
\label{fig:ee_limits} 
\end{center}
\end{figure}

\begin{figure}
\begin{center}
\epsfxsize=\textwidth
\epsfbox[0 150 595 742]{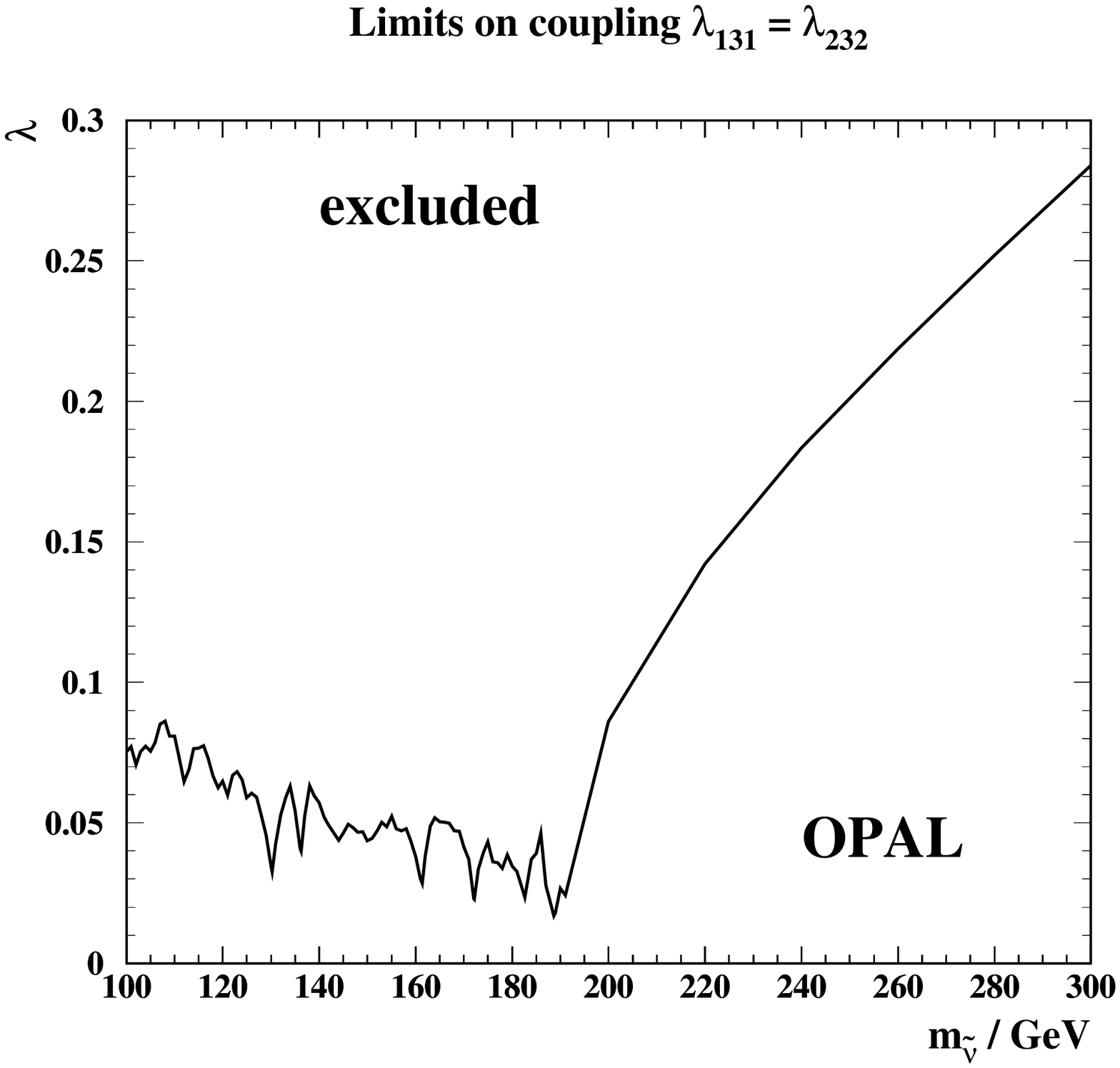}
\caption{95\% confidence exclusion limit on $\protect\lambda_{131} = 
 \lambda_{232}$ as a function of sneutrino mass $\protect m_{\snu}$, derived 
 from \mumu\ $s'$ distributions. The region above the solid
 line is excluded.
}
\label{fig:mu_limits} 
\end{center}
\end{figure}

The limits on $\lambda_{131}$ derived from 
the \epem\ data are shown in Fig.~\ref{fig:ee_limits}. They are in the range 
\mbox{0.01 -- 0.12} for $100 < \protect m_{\snu} < 200$~GeV. For masses
above 200~GeV there is no $s$-channel contribution, and the limits
rise to 0.16 at 300~GeV. Figure~\ref{fig:mu_limits} shows limits on 
$\lambda_{131} = \lambda_{232}$ derived from the \mumu\ data. These are in 
the range 0.02 -- 0.08 for $100 < \protect m_{\snu} < 200$~GeV, rising to 
0.28 at 300~GeV. The fine structure in the region 
$\protect m_{\snu} <$ 200~GeV results from fluctuations in 
the $s'$ distributions. 
These limits are valid for sneutrino widths of 1~GeV or less.
The inclusion of the 189 GeV data has significantly improved the limits
for masses near 189~GeV.

Direct searches~\cite{bib:ALEPH-rpv} for sneutrinos with $R$-parity 
violating couplings can exclude a $\snu_e$ with mass 
less than 80~GeV and a $\snu_\mu$ with mass less than 58~GeV. Our results
place limits on the couplings for masses above 100~GeV.

\subsection{Gravitational Interaction in Extra Dimensions}  \label{sec:gravity}
In nature there are two fundamental scales which differ by
many orders of magnitude, the ratio between the Planck scale
($M_{\rm Pl}\sim 10^{18-19}$~GeV)
and the electroweak scale ($M_{\rm EW}\sim 10^{2-3}$~GeV) being about 
$10^{16}$. 
The lack of explanation of this fact is known as the `hierarchy problem'.
Recently it has been pointed out that the quantum-gravity scale
could be as low as the electroweak scale with gravitons propagating 
in a compactified higher dimensional space~\cite{bib:ADD}, while other
Standard Model particles are confined to the usual $3 + 1$ space-time
dimensions.
According to this theory the Planck mass in $D=n+4$ dimensions
($M_{\rm D}$) is chosen to be the electroweak scale, so that the hierarchy 
problem is solved by definition. The Planck mass in the usual 4
space-time dimensions is given by
\begin{eqnarray}\label{eq-Mpl}
M_{\rm Pl}^2 =  R^n M_{\rm D}^{n+2}.
\end{eqnarray}   
where $R$ is the compactification radius of the extra dimensions.

Gravitons may contribute to two-fermion production via the process
$\epem \rightarrow {\rm G^*} \rightarrow \ffbar$. Although the contribution
from a single graviton state is very small compared with the Standard Model
contribution, the very large number of possible excitation modes in
the extra dimensions might lead to a measurable effect~\cite{bib:Giudice}.

The phenomenology of virtual graviton exchange processes in the context of 
collider experiments is described in~\cite{bib:Giudice,bib:Hewett,bib:Peskin,
bib:Rizzo}.
The differential cross-section 
for the production of fermion pairs 
with the inclusion of virtual graviton exchange 
can be written generally as 
\begin{eqnarray}\label{eq-grav}
\frac{{\rm d}\sigma}{{\rm d}\!\cos\theta} = A(\ct) 
+ B(\ct) \left[\frac{\lambda}{M_{\rm s}^4}\right] 
+ C(\ct) \left[\frac{\lambda}{M_{\rm s}^4}\right]^2,
\end{eqnarray}
where $\theta$
is the production polar angle of the fermion
with respect to the e$^-$ beam direction and
$M_{\rm s}$ is a mass scale parameter of the order of $M_{\rm D}$.
This parametrisation is taken from~\cite{bib:Hewett} and
we consistently use it in this paper.
The exact definition of the scale parameter can be found 
in~\cite{bib:Giudice}\footnote{Note that in the reference by 
Giudice $et~al.$~\cite{bib:Giudice} a different notation is used. 
The scale factor $\Lambda_T$ 
in the paper by Giudice $et~al.$ is
defined to be $\Lambda_T^4 = \frac{\pi}{2}\frac{M_{\rm s}^4}{|\lambda|}$.
If the ultra-violet cut-off for graviton exchange is taken as $M_{\rm D}$,
$M_{\rm s}$ is essentially the same as $M_{\rm D}$ except for an 
${\cal O}(1)$ factor.}. 
The parameter $\lambda$ is of ${\cal O}(1)$ and cannot be explicitly
calculated without knowledge of the full quantum 
gravity theory~\cite{bib:Hewett}. 
In contrast to graviton production, $\epem\rightarrow {\rm G}\gamma$, 
the dependence of the fermion-pair cross-section on the number of extra 
dimensions is weak and is included in $\lambda$.                     
Here we consider the cases $\lambda = +1$ and $\lambda = -1$. 

Although the functional form of the new interaction is similar to that of the 
contact interaction~\cite{bib:Eichten}, the differential cross-section 
for the new interaction includes terms proportional 
to $\cos^3 \theta$ or $\cos^4 \theta$.
The mass scale dependence of the 
amplitude of the new interaction is 1/(mass scale)$^4$, 
whereas that of the ordinary contact 
interaction is 1/(mass scale)$^2$. 

The first term in Equation~\eqref{eq-grav} is the Standard Model prediction,
the second term is the interference term and the 
third is the new interaction term.
The coefficients in the above expression are given in~\cite{bib:Hewett}.

We have analysed the angular distributions of non-radiative 
muon and tau pair events at 189~GeV, together with the
distributions at 183~GeV presented in~\cite{bib:OPAL-SM183}.
To obtain a lower limit on $M_{\rm s}$, we performed a binned 
maximum likelihood fit to angular distributions at the two
centre-of-mass energies simultaneously, in a similar manner to the
contact interaction analysis in Section~\ref{sec:ci}.
In order to fit the differential cross-section Equation~\eqref{eq-grav} to the
data,
a first-order photon radiation correction~\cite{bib:BK} was applied to the 
terms $B$ and $C$. ZFITTER was used to calculate the Standard Model term $A$.
The theoretical cross-section as a function of 
$\varepsilon\equiv\lambda/M_{\rm s}^4$ was then converted to the expected 
number of events in each of the $\cos\theta$ bins, taking into account the 
event selection efficiency, background, feedthrough of low $s^{\prime}$ 
events and the effect of interference between initial- and 
final-state photon radiation.
The likelihood was calculated from the Poisson probability for the 
observed number of events.
Additional Gaussian smearing was taken into account 
in the likelihood in order to
allow the overall normalization error to vary within the systematic errors
discussed in Section~\ref{sec:data}. We assigned a theoretical error
of 0.6\% to the Standard Model prediction of ZFITTER, as discussed in
Section~\ref{sec:sm}.

We derived the 95\% confidence level lower limits on $M_{\rm s}$ from the 
values of $\varepsilon$ corresponding to an increase in the negative log 
likelihood of 1.92 with
respect to the minimum found in the $\varepsilon$ region considered.
As $M_{\rm s}^4$ must be positive, the physically allowed region is
$\varepsilon > 0$ for $\lambda=+1$ and $\varepsilon < 0$ for $\lambda=-1$. 

\begin{figure}
\begin{center}
\epsfxsize=\textwidth
\epsfbox{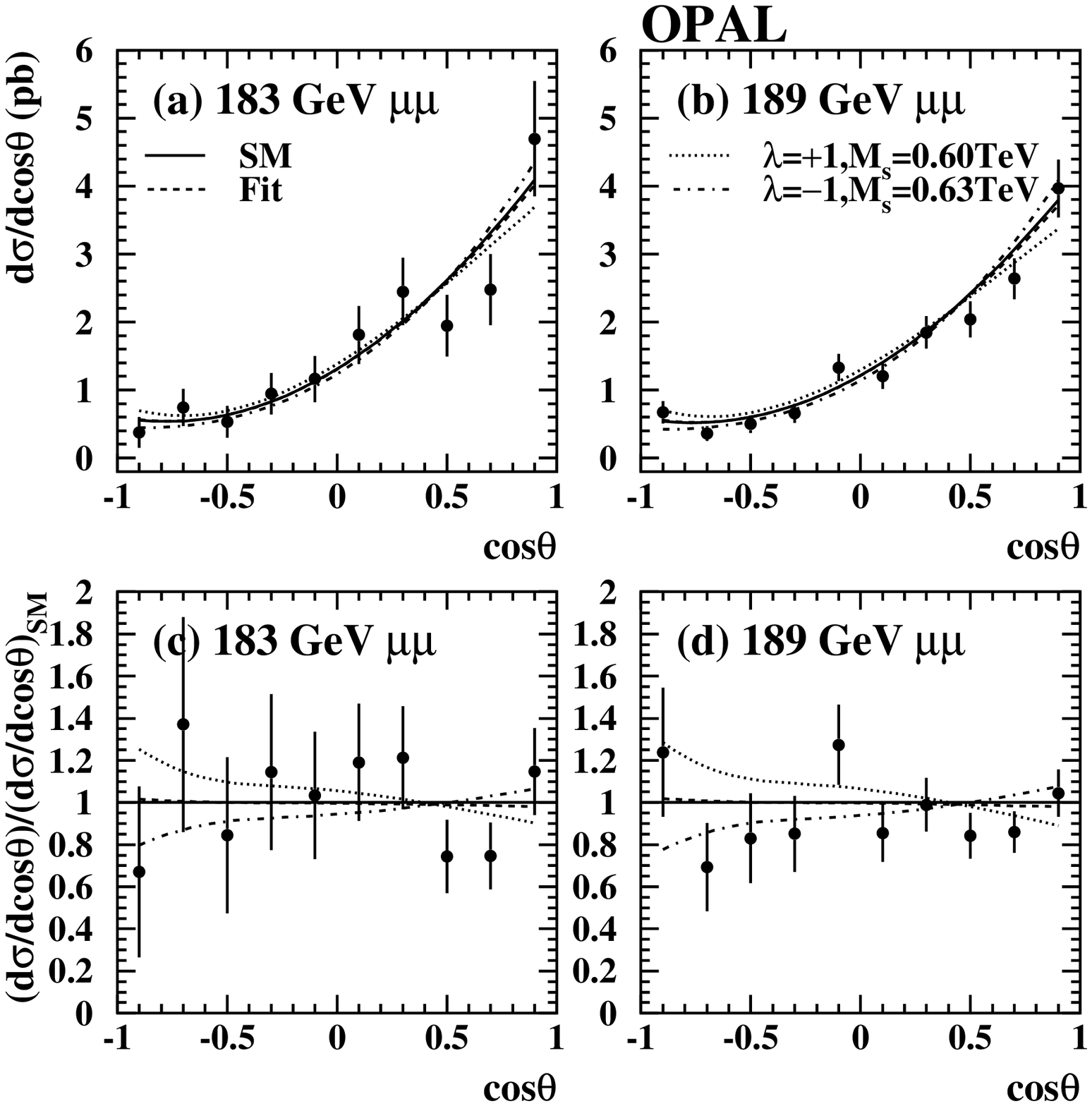}
\caption{The differential cross-sections for muon pairs at 
         (a) $\protect \sqrt{s} = 183$~GeV and 
         (b) $\protect \sqrt{s} = 189$~GeV.
         The curves show the Standard Model prediction (solid line), 
         the best fit with the new gravitational interaction (dashed line), 
         and the distributions corresponding to the 95\% confidence level
         limits on $M_{\rm s}$ with $\lambda = +1$ (dotted line) and
         $\lambda = -1$ (dot-dashed line). 
         Note that these curves correspond to the simultaneous fit of
         183~GeV and 189~GeV data.
         The ratios of data to Standard Model prediction for 
         (c) 183~GeV muon pairs and (d) 189~GeV muon pairs are also
         shown.
}
\label{fig:grav_mu}
\end{center}
\end{figure}

\begin{figure} 
\begin{center}
\epsfxsize=\textwidth
\epsfbox{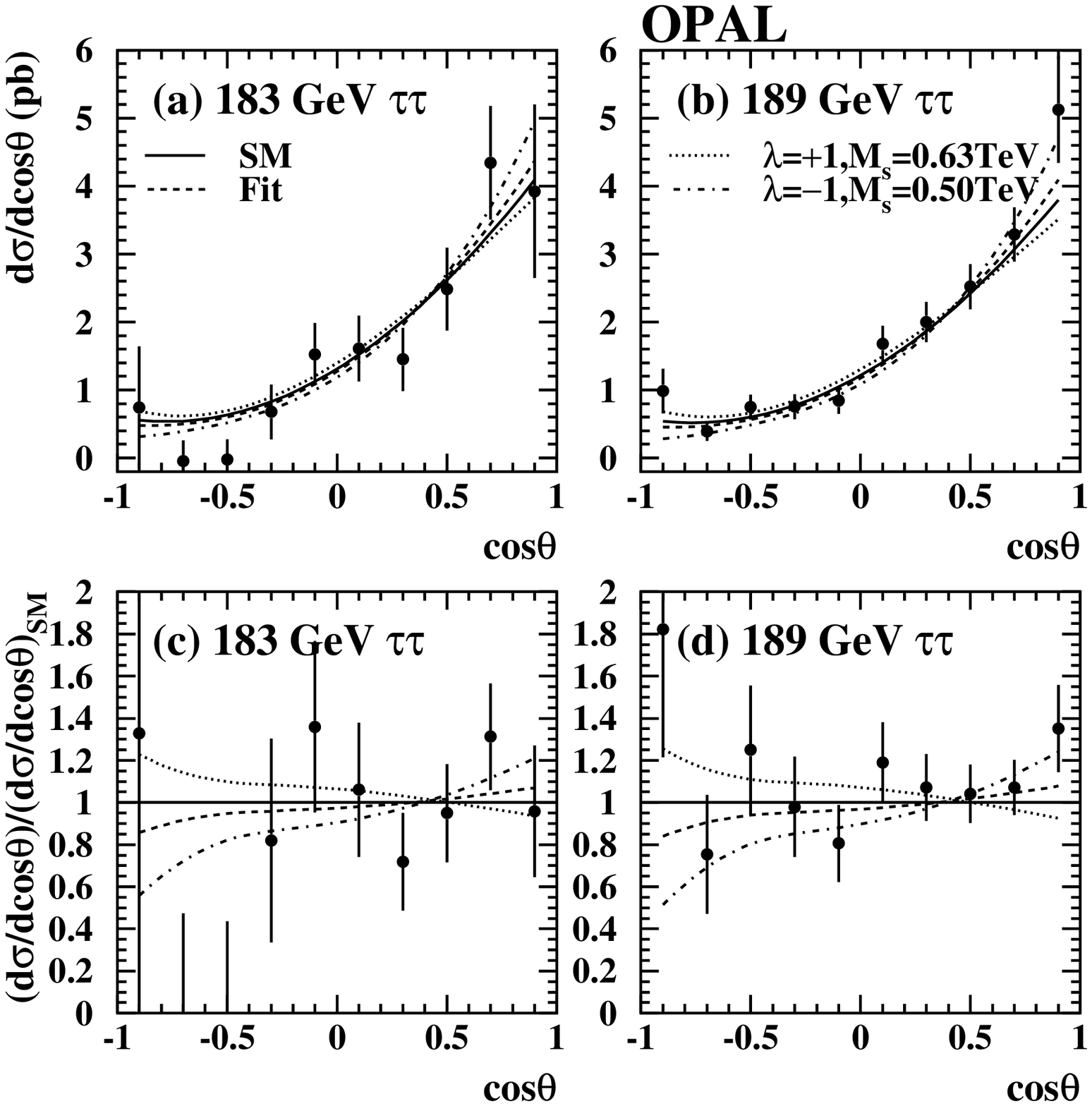}
\caption{The differential cross-sections for tau pairs at 
         (a) $\protect \sqrt{s} = 183$~GeV and 
         (b) $\protect \sqrt{s} = 189$~GeV.
         The curves show the Standard Model prediction (solid line), 
         the best fit with the new gravitational interaction (dashed line), 
         and the distributions corresponding to the 95\% confidence level
         limits on $M_{\rm s}$ with $\lambda = +1$ (dotted line) and
         $\lambda = -1$ (dot-dashed line). 
         Note that these curves correspond to the simultaneous fit of
         183~GeV and 189~GeV data.
         The ratios of data to Standard Model prediction for 
         (c) 183~GeV tau pairs and (d) 189~GeV tau pairs are also
         shown.
}
\label{fig:grav_tau}
\end{center}
\end{figure}

The 95\% confidence level lower limits on $M_{\rm s}$ derived from the 
muon pairs, from the tau pairs, and from a simultaneous fit to the muon
and tau pairs, are given in Table~\ref{tab:grav}. They are in the
range 0.50--0.68~TeV. The results of the fit to the muon pairs are
displayed in Fig.~\ref{fig:grav_mu}, those of the fit to the tau
pairs in Fig.~\ref{fig:grav_tau}. In each case we show the measured angular
distributions and the ratio of the measurements to the Standard Model
predictions, together with curves representing the Standard Model
prediction, the best fit, and the distributions corresponding to
the 95\% confidence level limits on $M_{\rm s}$.

Limits on the gravitational interaction in extra dimensions have also been 
derived from OPAL measurements of photonic final states~\cite{bib:OPAL-gg189}.

\section{Conclusions}    \label{sec:sum}
We have presented new measurements of cross-sections and asymmetries for
hadron and lepton pair production in $\epem$ collisions at a centre-of-mass
energy of 189~GeV. The results, for both inclusive fermion-pair production 
and for non-radiative events, are in good agreement with Standard Model 
expectations. From these and earlier measurements we derive a value for the 
electromagnetic coupling constant 
$1/\alphaem (181.94~\mathrm{GeV}) = 126.8^{+3.0}_{-2.7}$.

The measurements have been used to improve existing limits on new
physics. In the context of a four-fermion contact interaction we
have improved the limits on the energy scale $\Lambda$ from typically
2--10~TeV to 3--13~TeV, assuming $g^2/4\pi = 1$. 
We have also presented limits on new particles such as sneutrinos in
supersymmetric theories with $R$-parity violation which couple to
leptons. Sensitivity to sneutrino masses between the centre-of-mass
energy points of LEP has been improved by using a complete scan of the
$s'$ distribution for processes involving an $s$-channel diagram. In
these cases, limits on the couplings in the range 0.01 -- 0.1 are obtained 
for $100 < m < 200$~GeV.

In a search for the possible effects of gravitons propagating in extra
dimensions, we have obtained lower limits on the effective Planck scale
in the space with extra dimensions in the range 0.50--0.68~TeV.

\section*{Acknowledgements}
We particularly wish to thank the SL Division for the efficient operation
of the LEP accelerator at all energies
 and for their continuing close cooperation with
our experimental group.  We thank our colleagues from CEA, DAPNIA/SPP,
CE-Saclay for their efforts over the years on the time-of-flight and trigger
systems which we continue to use.  In addition to the support staff at our own
institutions we are pleased to acknowledge the  \\
Department of Energy, USA, \\
National Science Foundation, USA, \\
Particle Physics and Astronomy Research Council, UK, \\
Natural Sciences and Engineering Research Council, Canada, \\
Israel Science Foundation, administered by the Israel
Academy of Science and Humanities, \\
Minerva Gesellschaft, \\
Benoziyo Center for High Energy Physics,\\
Japanese Ministry of Education, Science and Culture (the
Monbusho) and a grant under the Monbusho International
Science Research Program,\\
Japanese Society for the Promotion of Science (JSPS),\\
German Israeli Bi-national Science Foundation (GIF), \\
Bundesministerium f\"ur Bildung, Wissenschaft,
Forschung und Technologie, Germany, \\
National Research Council of Canada, \\
Research Corporation, USA,\\
Hungarian Foundation for Scientific Research, OTKA T-029328, 
T023793 and OTKA F-023259.\\

%
\clearpage
\begin{table}[htbp]
\centering
\begin{tabular}{|ll|c|c|c|}
\hline
\hline
\multicolumn{5}{|c|}{\bf Efficiencies and backgrounds at $\sqrt{s}$ = 189~GeV}
\\
\hline
Channel  & &Efficiency (\%)  &Background (pb) &Feedthrough (pb) \\ 
\hline
\qqbar X & &90.3$\pm$0.5 &4.2$\pm$0.9 &-- \\
\hline
\qqbar &$s'/s>0.01$ &87.3$\pm$0.5 &6.5$\pm$0.9 &-- \\
       &$s'/s>0.7225$ & 87.7$\pm$0.7 &1.55$\pm$0.09 &1.1$\pm$0.1 \\ 
\hline
\Pep\Pem &$\absct<0.9$, $\thacol<170\degree$  &97.8$\pm$0.6 
                                              &1.57$\pm$0.09   &-- \\
         &$\absctem<0.7$, $\thacol<10\degree$ &98.9$\pm$0.4 
                                               &0.25$\pm$0.03  &-- \\
         &$\absct<0.96$, $\thacol<10\degree$  &98.5$\pm$0.4 
                                               &10.4$\pm$0.4   &-- \\
\hline
\Pgmp\Pgmm &$s'/s>0.01$ &75.4$\pm$0.8 &0.39$\pm$0.11 &-- \\
           &$s'/s>0.7225$ &88.7$\pm$0.9 &0.07$\pm$0.03 &0.060$\pm$0.003 \\
\hline
\Pgtp\Pgtm &$s'/s>0.01$ &40.2$\pm$1.0 &0.54$\pm$0.10 &-- \\
           &$s'/s>0.7225$ &58.3$\pm$1.5 &0.17$\pm$0.03 &0.072$\pm$0.003 \\
\hline
\hline
\end{tabular}
\caption[]{
  Efficiency of selection cuts, background and feedthrough of events
  with lower $s'$ into the non-radiative samples for each channel at 
  189~GeV. The errors include Monte Carlo statistics and systematic 
  effects. In the case of electron pairs, the efficiencies are
  effective values including the efficiency of selection cuts for events
  within the acceptance region and the effect of acceptance corrections.  
  An acceptance of $\absct<0.9$ (or 0.96) means that both electron and
  positron must satisfy this cut, whereas $\absctem<0.7$ means
  that only the electron need do so.
}
\label{tab:eff}
\end{table}

\begin{table}[htbp]
\centering
\begin{tabular}{|ll|c|r|r@{$\pm$}l@{$\pm$}l|c|}
\hline
\hline
\multicolumn{8}{|c|}{\bf Cross-sections at $\sqrt{s}$ = 189~GeV} \\
\hline
Channel   &       &$\int{\cal L}\mathrm{d}t$ (pb$^{-1}$) &  Events & 
    \multicolumn{3}{c|}{$\sigma$ (pb)} &  $\sigma^{\mathrm{SM}}$ (pb) \\
\hline
\qqbar X  & &185.6 &20025  & 114.8&0.9&1.2   & 114.3   \\
\hline
\qqbar &$s'/s>0.01$   &185.6 &17228 & 99.5&0.8&1.2      & 98.8  \\
       &$s'/s>0.7225$ &      & 4072 & 22.10&0.37&0.24   & 22.16 \\
\hline
\Pep\Pem &$\absct<0.9$, $\thacol<170\degree$  &185.8 &20487 &
                                              111.2&0.8&0.7 & 112.0 \\
         &$\absctem<0.7$, $\thacol<10\degree$ &      & 3735 &
                                            20.08&0.33&0.10 & 20.39 \\
         &$\absct<0.96$, $\thacol<10\degree$  &      & 57685 &
                                              304.6&1.3&1.4 & 311.6 \\
\hline
\Pgmp\Pgmm &$s'/s>0.01$   &180.0 &1129 &7.77&0.23&0.18  & 7.76 \\
           &$s'/s>0.7225$ &      & 527 &3.11&0.14&0.06  & 3.21 \\
\hline
\Pgtp\Pgtm &$s'/s>0.01$   &180.6 & 730 &8.67&0.32&0.34  & 7.75 \\
           &$s'/s>0.7225$ &      & 420 &3.54&0.17&0.11  & 3.21 \\
\hline
\hline
\end{tabular}
\caption[]{
  Integrated luminosity used in the analysis, numbers of selected events 
  and measured cross-sections at $\sqrt{s}$=188.63~GeV. 
  For the cross-sections, the first error shown is statistical, the second
  systematic.  As in~\cite{bib:OPAL-SM183,bib:OPAL-SM172}, the cross-sections
  for hadrons, \Pgmp\Pgmm\ and \Pgtp\Pgtm\ 
  are defined to cover phase-space up to the limit imposed by the $s'/s$ cut, 
  with $\sqrt{s'}$ defined as the invariant mass
  of the outgoing two-fermion system {\em before} final-state photon
  radiation. The contribution of interference between initial-
  and final-state radiation has been removed.
  The last column shows the Standard Model cross-section predictions from
  ZFITTER~\cite{bib:zfitter} (hadrons, \Pgmp\Pgmm, \Pgtp\Pgtm)
  and  ALIBABA~\cite{bib:alibaba} (\Pep\Pem).
}
\label{tab:xsec}
\end{table}

\begin{table}[htbp]
\centering
\begin{tabular}{|ll|r|r|c|c|}
\hline
\hline
\multicolumn{6}{|c|}{\bf Asymmetries at $\sqrt{s}$ = 189~GeV} \\
\hline
&&$N_{\mathrm{F}}$ &$N_{\mathrm{B}}$ &$\AFB$ &$\AFBSM$ \\ 
\hline
\epem    &$\absctem < 0.7$ & 3318 &349 &0.814$\pm$0.010$\pm$0.005 &0.813 \\
         &and $\thacol < 10^\circ$ & & & &                \\ 
\hline
\mumu    &$s'/s > 0.01$    &682    &372   &0.253$\pm$0.031$\pm$0.003 &0.277 \\
         &$s'/s > 0.7225$  &370    &115.5 &0.532$\pm$0.042$\pm$0.007 &0.566 \\
\hline
\tautau  &$s'/s > 0.01$    & 485   &217   &0.315$\pm$0.042$\pm$0.003 &0.278 \\
         &$s'/s > 0.7225$  & 316.5 & 89.5 &0.606$\pm$0.048$\pm$0.007 &0.566 \\
\hline
Combined &$s'/s > 0.01$    &           &  &0.277$\pm$0.025$\pm$0.002 &0.277 \\
\mumu\ and \tautau &$s'/s > 0.7225$  & &  &0.563$\pm$0.031$\pm$0.005 &0.566 \\
\hline
\hline
\end{tabular}
\caption[]{The numbers of forward ($N_{\mathrm{F}}$) and backward
  ($N_{\mathrm{B}}$) events and measured asymmetry values at 188.63~GeV. The
  measured asymmetry values include corrections for background and efficiency,
  and in the case of muons and taus are corrected to the full solid angle.
  The first error is statistical and the second systematic.
  The asymmetries for
  \Pgmp\Pgmm, \Pgtp\Pgtm\ and for the combined \mumu\ and \tautau\ are
  shown after the correction for interference between
  initial- and final-state radiation.
  The final column shows the Standard Model predictions of ALIBABA for
  \epem\ and ZFITTER for the other final states.
}
\label{tab:afb}
\end{table}

\begin{table}[htbp]
\begin{center}
\begin{tabular} {|c|c|}
\hline
\hline
\multicolumn{2}{|c|}{\bf \boldmath Hadrons at $\sqrt{s}$ = 189~GeV} \\
\hline
$\absct$   &\multicolumn{1}{c|}{$\dsdabscc$ (pb)} \\
\hline
$ [0.0,0.1]$ &17.5$\pm$1.0 \\
$ [0.1,0.2]$ &17.7$\pm$1.1 \\
$ [0.2,0.3]$ &16.8$\pm$1.0 \\
$ [0.3,0.4]$ &18.2$\pm$1.1 \\
$ [0.4,0.5]$ &18.8$\pm$1.1 \\
$ [0.5,0.6]$ &21.6$\pm$1.2 \\
$ [0.6,0.7]$ &24.2$\pm$1.2 \\
$ [0.7,0.8]$ &26.0$\pm$1.3 \\
$ [0.8,0.9]$ &27.7$\pm$1.3 \\
$ [0.9,1.0]$ &31.4$\pm$1.7 \\
\hline
\hline
\end{tabular}     
\caption{
Differential cross-section for $\qqbar$ production, for $s'/s > 0.7225$.  
The values are corrected to no interference between initial- and final-state 
radiation as in~\cite{bib:OPAL-SM172}. Errors include statistical and 
systematic effects combined, with the former dominant.
}
\label{tab:mh_angdis}
\end{center}  
\end{table}

\begin{table}[htbp]
\begin{center}
\small
\begin{tabular} {|c|r@{$\pm$}l|}
\hline
\hline
\multicolumn{3}{|c|}{\bf \boldmath $\epem$ at $\sqrt{s}$ = 189~GeV} \\
\hline
$\ct$  &\multicolumn{2}{c|}{$\dsdcc$ (pb)} \\
\hline
$[-0.9,-0.7]$            & 1.4&0.2 \\
$[-0.7,-0.5]$            & 2.0&0.2 \\
$[-0.5,-0.3]$            & 2.4&0.3 \\
$[-0.3,-0.1]$            & 3.0&0.3 \\
$[-0.1,\;\;\;0.1]$       & 4.3&0.3 \\
$[\;\;\;0.1,\;\;\;0.3]$  & 8.3&0.5 \\
$[\;\;\;0.3,\;\;\;0.5]$  &19.3&0.7 \\
$[\;\;\;0.5,\;\;\;0.7]$  &61.4&1.4 \\
$[\;\;\;0.7,\;\;\;0.9]$  &415 &5   \\
\hline
\hline
\end{tabular}     
\caption{
Differential cross-section for electron pair production for
$\thacol < 10\degree$. Errors include statistical 
and systematic effects combined.
}
\label{tab:ee_angdis}
\end{center}  
\end{table}

\begin{table}[p]
\begin{center}
\small
\begin{tabular} {|c|r@{$\pm$}l|}
\hline
\hline
\multicolumn{3}{|c|}{\bf \boldmath $\mumu$ at $\sqrt{s}$ = 189~GeV} \\
\hline
$\ct$  &\multicolumn{2}{c|}{$\dsdcc$ (pb)} \\
\hline
$[-1.0,-0.8]$ &$0.67$&$^{0.21}_{0.17}$ \\
$[-0.8,-0.6]$ &$0.36$&$^{0.14}_{0.11}$ \\
$[-0.6,-0.4]$ &$0.50$&$^{0.16}_{0.13}$ \\
$[-0.4,-0.2]$ &$0.66$&$0.14$ \\
$[-0.2,\;\;\;0.0]$  &$1.33$&$0.20$ \\
$[\;\;\;0.0,\;\;\;0.2]$ &$1.20$&$0.19$ \\
$[\;\;\;0.2,\;\;\;0.4]$ &$1.85$&$0.24$ \\
$[\;\;\;0.4,\;\;\;0.6]$ &$2.04$&$0.27$ \\
$[\;\;\;0.6,\;\;\;0.8]$ &$2.64$&$0.30$ \\
$[\;\;\;0.8,\;\;\;1.0]$ &$3.96$&$0.43$ \\
\hline
\hline
\multicolumn{3}{|c|}{\bf \boldmath $\tautau$ at $\sqrt{s}$ = 189~GeV} \\
\hline
$\ct$  &\multicolumn{2}{c|}{$\dsdcc$ (pb)} \\
\hline
$[-1.0,-0.8]$ &$0.99$&$^{0.44}_{0.33}$ \\
$[-0.8,-0.6]$ &$0.39$&$^{0.19}_{0.15}$ \\
$[-0.6,-0.4]$ &$0.75$&$0.18$ \\
$[-0.4,-0.2]$ &$0.76$&$0.18$ \\
$[-0.2,\;\;\;0.0]$ &$0.84$&$0.19$ \\
$[\;\;\;0.0,\;\;\;0.2]$ &$1.68$&$0.27$ \\
$[\;\;\;0.2,\;\;\;0.4]$ &$2.00$&$0.30$ \\
$[\;\;\;0.4,\;\;\;0.6]$ &$2.52$&$0.33$ \\
$[\;\;\;0.6,\;\;\;0.8]$ &$3.29$&$0.40$ \\
$[\;\;\;0.8,\;\;\;1.0]$ &$5.1$&$0.8$ \\
\hline
\hline
\end{tabular}     
\caption{
Differential cross-sections for \mumu\ and \tautau\ pair production. 
The values are for $s'/s > 0.7225$ and are corrected to no interference between 
initial- and final-state radiation. Errors include statistical 
and systematic effects combined, with the former dominant.
}
\label{tab:mu_angdis}
\end{center}  
\end{table}

\begin{table}[htbp]
\centering
\begin{tabular}{|l|l|l|}
\hline
\hline
\multicolumn{3}{|c|}{\bf Interference Corrections at $\sqrt{s}$ = 189~GeV} \\
\hline
          &\boldmath $s'/s>0.01$  &\boldmath $s'/s>0.7225$  \\
\hline
$\Delta\sigma/\sigma_{\rm SM}$(had) (\%)  
             & $+0.05\pm0.00\pm0.05$  & $+0.4\pm0.1\pm0.2$ \\
$\Delta\sigma/\sigma_{\rm SM}(\mu\mu)$ (\%)
             & $-0.42\pm0.05$         & $-1.4\pm0.4$       \\
$\Delta\sigma/\sigma_{\rm SM}(\tau\tau)$ (\%)
             & $-0.48\pm0.06$         & $-1.2\pm0.3$       \\
$\Delta A_{\rm FB}(\mu\mu)$
             & $-0.0054\pm0.0007$     & $-0.016\pm0.004$   \\
$\Delta A_{\rm FB}(\tau\tau)$
             & $-0.0053\pm0.0009$     & $-0.012\pm0.003$   \\
$\Delta A_{\rm FB}$(combined)
             & $-0.0056\pm0.0006$     & $-0.014\pm0.003$   \\
\hline
\hline
\end{tabular}
\caption[]{Corrections $\Delta\sigma$ and $\Delta A_{\rm FB}$
 which have been applied to the measured cross-sections and 
 asymmetries in order to remove the contribution from interference
 between initial- and final-state radiation. Cross-section corrections are
 expressed as a fraction of the expected Standard Model cross-section,
 while asymmetry corrections are given as absolute numbers, and
 depend on the observed asymmetry. The first error reflects the uncertainty 
 from  modelling the selection efficiency for the interference cross-section, 
 and is very small for hadrons because the efficiency is large and
 depends only weakly on \ct. The second error is our estimate of 
 possible additional QCD corrections for the hadrons~\cite{bib:OPAL-SM172}.
}
\label{tab:ifsr}
\end{table}

\begin{table}[htbp]
\begin{center}
\begin{tabular}{|l|c|c|}
\hline
                  &\boldmath $s'/s>0.01$  &\boldmath $s'/s>0.7225$  \\
\hline
MC statistics (efficiency) &0.05             &0.14 \\
MC statistics (background) &0.27             &0.14 \\
ISR modelling              &0.45             &0.48 \\
Fragmentation modelling    &0.34             &0.32 \\
Detector effects           &0.19             &0.42 \\
$s'$ determination         &--               &0.30 \\
WW rejection cuts          &0.11             &0.52 \\
WW background              &0.15             &0.39 \\
Background                 &0.98             &0.09 \\
Interference               &0.05             &0.22 \\
Luminosity                 &0.21             &0.21 \\
\hline
Total                      &1.2              &1.1  \\
\hline
\end{tabular}
\end{center}
\caption{Systematic errors, in \%, on the hadronic cross-section 
         measurements.
        }
\label{tab:mh_syserr}
\end{table}

\begin{table}
\begin{center}
\begin{tabular}{|l|c|c|c|}
\hline
       &$\absctepem < 0.9$  &$\absctem < 0.7$  &$\absctepem < 0.96$ \\
       &$\thacol < 170^{\circ}$ &$\thacol < 10^{\circ}$ &$\thacol < 10^{\circ}$
       \\
\hline
MC statistics                       &0.05   &0.10   &0.05   \\
Four-fermion contribution           &0.03   &0.06   &0.01   \\
Multiplicity cuts                   &0.12   &0.05   &0.03   \\
Calorimeter energy scale/resolution &0.01   &0.01   &0.10   \\
Track requirements                  &0.42   &0.38   &--     \\
Acceptance correction               &0.37   &0.19   &0.37   \\
Background                          &0.08   &0.15   &0.12   \\
Luminosity                          &0.21   &0.21   &0.21   \\ \hline
Total                               &0.6    &0.5    &0.5    \\
\hline
\end{tabular}
\end{center}
\caption{Systematic errors, in \%, on the electron pair cross-section 
         measurements. 
        }
\label{tab:ee_syserr}
\end{table}

\begin{table}
\begin{center}
\begin{tabular}{|l|c|c|}
\hline
                  &\boldmath $s'/s>0.01$  &\boldmath $s'/s>0.7225$  \\
\hline
MC statistics (efficiency)  &0.2          &0.2  \\
MC statistics (background)  &0.2          &0.2  \\
MC statistics (feedthrough) &--           &0.1  \\
Efficiency                  &1.0          &1.0  \\
Cosmic background           &0.5          &1.0  \\
Other background            &1.9          &1.1  \\
Feedthrough                 &--           &0.1  \\
Interference                &$<$0.1       &0.4  \\
Luminosity                  &0.2          &0.2  \\
\hline
Total                       &2.3          &1.8  \\
\hline
\end{tabular}
\end{center}
\caption{Systematic errors, in \%, on the muon pair cross-section 
         measurements.
        }
\label{tab:mu_syserr}
\end{table}

\begin{table}
\begin{center}
\begin{tabular}{|l|c|c|}
\hline
                  &\boldmath $s'/s>0.01$  &\boldmath $s'/s>0.7225$  \\
\hline
MC statistics (efficiency)  &0.4          &0.5  \\
MC statistics (background)  &0.6          &0.5  \\
MC statistics (feedthrough) &--           &0.1  \\
Efficiency                  &2.5          &2.5  \\
Background                  &2.9          &1.5  \\
Feedthrough                 &--           &0.1  \\
Interference                &0.1          &0.3  \\
Luminosity                  &0.2          &0.2  \\
\hline
Total                       &3.9          &3.0  \\
\hline
\end{tabular}
\end{center}
\caption{Systematic errors, in \%, on the tau pair cross-section 
         measurements.
        }
\label{tab:tau_syserr}
\end{table}

\begin{table}[htbp]
\centering
\begin{tabular}{|c||c|c||c|c|}
\hline
\hline
      &\multicolumn{2}{c||}{Fit} &\multicolumn{2}{c|}{Standard Model} \\
\hline
$\sqrt{s}$ (GeV) &$1/\alphaem$ &$\chi^2$/d.o.f. &$1/\alphaem$ 
&$\chi^2$/d.o.f. \\
\hline
188.63  &126.2$^{+3.7}_{-3.2}$  & 3.0/3  & 127.8  & 3.2/4 \\
\hline
181.94  &126.8$^{+3.0}_{-2.7}$  &15.0/23 & 127.9  &15.2/24 \\
\hline
\hline
\end{tabular}
\caption[]{Results of fits for $\alphaem$. The first row shows the fit
           to data at 188.63~GeV, the second row the combined fit to 
           these data and measurements at 
           130--183~GeV~\cite{bib:OPAL-SM183,bib:OPAL-SM172}. 
           For the combined fit, the value of \alphaem\ is quoted at the 
           centre-of-mass energy corresponding to the luminosity-weighted 
           average of $1/s$. 
           The errors on the fitted values of \alphaem\ arise from the 
           errors on the measurements; errors due to uncertainties in the 
           ZFITTER input parameters are negligible.            
           The Standard Model values of $1/\alphaem$, and
           the $\chi^2$ between the measurements and the Standard Model 
           predictions are also given for comparison.
}
\label{tab:alphaem1}
\end{table}

\begin{table}[htbp]
\begin{sideways}
\begin{minipage}[b]{\textheight}{\footnotesize
\begin{center}\begin{tabular}{|cc|c|c|c|c|c|c|c|c|c|}
\hline
Channel &      &  LL  &  RR  &  LR  &  RL  &  VV  &  AA  &  
               LL+RR  &LR+RL & $\overline{\cal O}_{\mathrm{DB}}$ \\
        &      &  \scriptsize{ $[\pm1,0,0,0]$}  & 
                  \scriptsize{ $[0,\pm1,0,0]$}  & 
                  \scriptsize{ $[0,0,\pm1,0]$}  &
                  \scriptsize{ $[0,0,0,\pm1]$}  & 
                  \scriptsize{ $[\pm1,\pm1,\pm1,\pm1]$} &
                  \scriptsize{ $[\pm1,\pm1,\mp1,\mp1]$} &
                  \scriptsize{ $[\pm1,\pm1,0,0]$} & 
                  \scriptsize{ $[0,0,\pm1,\pm1]$} &
       \scriptsize{ $[\pm\frac{1}{4},\pm1,\pm\frac{1}{2},\pm\frac{1}{2}]$} \\

 \hline
\epem    &$\epsz$& $ 0.016_{-0.024}^{+0.026}$ & $ 0.016_{-0.024}^{+0.026}$ & 
                   $ 0.008_{-0.013}^{+0.013}$ & $ 0.008_{-0.013}^{+0.013}$ & 
                   $ 0.003_{-0.005}^{+0.005}$ & $-0.002_{-0.008}^{+0.008}$ & 
                   $ 0.008_{-0.012}^{+0.012}$ & $ 0.004_{-0.007}^{+0.007}$ & 
                   $ 0.006_{-0.009}^{+0.009}$ \\
         &$\lamp$& 3.8 & 3.8 & 5.3 & 5.3 & 8.9 & 8.9 & 5.5 & 7.7 & 6.5 \\
         &$\lamm$& 5.6 & 5.5 & 7.5 & 7.5 &12.4 & 7.2 & 7.7 &10.4 & 9.1 \\
 \hline
\mumu    &$\epsz$& $-0.015_{-0.016}^{+0.015}$ & $-0.018_{-0.017}^{+0.018}$ & 
                   $-0.003_{-0.021}^{+0.018}$ & $-0.003_{-0.021}^{+0.018}$ & 
                   $-0.005_{-0.006}^{+0.006}$ & $-0.005_{-0.007}^{+0.007}$ & 
                   $-0.008_{-0.008}^{+0.008}$ & $-0.002_{-0.009}^{+0.010}$ & 
                   $-0.009_{-0.010}^{+0.010}$ \\
         &$\lamp$& 7.3 & 7.0 & 5.5 & 5.5 &11.3 &10.3 &10.0 & 7.6 & 8.6 \\
         &$\lamm$& 4.6 & 4.4 & 1.9 & 1.9 & 7.9 & 7.5 & 6.5 & 7.0 & 5.8 \\
 \hline
\tautau  &$\epsz$& $ 0.024_{-0.020}^{+0.021}$ & $ 0.027_{-0.023}^{+0.022}$ & 
                   $-0.239_{-0.031}^{+0.034}$ & $-0.239_{-0.031}^{+0.034}$ & 
                   $ 0.005_{-0.008}^{+0.008}$ & $ 0.011_{-0.009}^{+0.009}$ & 
                   $ 0.013_{-0.011}^{+0.011}$ & $-0.244_{-0.016}^{+0.016}$ & 
                   $ 0.010_{-0.014}^{+0.014}$ \\
         &$\lamp$& 3.9 & 3.8 & 5.0 & 5.0 & 6.9 & 5.9 & 5.3 & 6.8 & 5.1 \\
         &$\lamm$& 6.5 & 6.2 & 1.9 & 1.9 & 9.5 &10.0 & 9.1 & 1.9 & 7.2 \\
 \hline
$\lept$  &$\epsz$& $ 0.002_{-0.011}^{+0.011}$ & $ 0.003_{-0.012}^{+0.012}$ & 
                   $ 0.002_{-0.010}^{+0.010}$ & $ 0.002_{-0.010}^{+0.010}$ & 
                   $ 0.001_{-0.003}^{+0.003}$ & $ 0.000_{-0.004}^{+0.004}$ & 
                   $ 0.001_{-0.006}^{+0.006}$ & $ 0.001_{-0.005}^{+0.005}$ & 
                   $ 0.002_{-0.006}^{+0.006}$ \\
         &$\lamp$& 6.4 & 6.2 & 6.8 & 6.8 &11.5 &10.9 & 8.9 & 9.5 & 8.5 \\
         &$\lamm$& 7.2 & 7.0 & 7.7 & 7.7 &13.4 &10.6 &10.1 &10.9 & 9.9 \\
 \hline
\qqbar   &$\epsz$& $-0.064_{-0.026}^{+0.080}$ & $ 0.029_{-0.052}^{+0.036}$ & 
                   $ 0.008_{-0.041}^{+0.041}$ & $ 0.111_{-0.027}^{+0.018}$ & 
                   $ 0.021_{-0.031}^{+0.016}$ & $-0.037_{-0.013}^{+0.044}$ & 
                   $-0.012_{-0.025}^{+0.032}$ & $ 0.062_{-0.072}^{+0.015}$ & 
                   $ 0.070_{-0.079}^{+0.016}$ \\
         &$\lamp$& 5.5 & 3.5 & 3.8 & 2.7 & 4.7 & 8.1 & 5.4 & 3.4 & 3.2 \\
         &$\lamm$& 3.1 & 4.9 & 4.4 & 6.4 & 7.2 & 4.2 & 4.4 & 7.1 & 6.7 \\
 \hline
combined &$\epsz$& $ 0.001_{-0.010}^{+0.010}$ & $ 0.003_{-0.010}^{+0.010}$ & 
                   $ 0.003_{-0.010}^{+0.010}$ & $ 0.003_{-0.008}^{+0.009}$ & 
                   $ 0.001_{-0.003}^{+0.003}$ & $-0.001_{-0.004}^{+0.004}$ & 
                   $ 0.001_{-0.006}^{+0.006}$ & $ 0.002_{-0.005}^{+0.005}$ & 
                   $ 0.002_{-0.006}^{+0.006}$ \\
         &$\lamp$& 7.1 & 6.8 & 6.8 & 6.9 &11.4 &11.6 & 9.0 & 9.5 & 8.5 \\
         &$\lamm$& 7.2 & 7.4 & 7.8 & 8.6 &13.7 &10.5 &10.0 &11.4 &10.5 \\
 \hline
\uubar   &$\epsz$& $ 0.005_{-0.017}^{+0.018}$ & $ 0.009_{-0.025}^{+0.027}$ & 
                   $ 0.144_{-0.182}^{+0.056}$ & $ 0.000_{-0.076}^{+0.103}$ & 
                   $ 0.002_{-0.008}^{+0.009}$ & $ 0.004_{-0.012}^{+0.014}$ & 
                   $ 0.003_{-0.010}^{+0.011}$ & $ 0.012_{-0.043}^{+0.113}$ & 
                   $ 0.004_{-0.016}^{+0.019}$ \\
         &$\lamp$& 4.9 & 1.5 & 2.1 & 2.6 & 7.0 & 5.4 & 6.4 & 2.5 & 4.9 \\
         &$\lamm$& 6.1 & 5.1 & 3.7 & 2.9 & 8.6 & 7.3 & 7.8 & 4.2 & 6.1 \\
 \hline
\ddbar   &$\epsz$& $-0.007_{-0.021}^{+0.020}$ & $-0.191_{-0.050}^{+0.221}$ & 
                   $-0.042_{-0.095}^{+0.098}$ & $ 0.152_{-0.184}^{+0.058}$ & 
                   $-0.005_{-0.020}^{+0.015}$ & $-0.004_{-0.014}^{+0.012}$ & 
                   $-0.004_{-0.015}^{+0.014}$ & $ 0.040_{-0.078}^{+0.051}$ & 
                   $-0.123_{-0.044}^{+0.152}$ \\
         &$\lamp$& 5.7 & 4.0 & 3.2 & 2.0 & 6.5 & 7.3 & 6.7 & 2.9 & 4.2 \\
         &$\lamm$& 4.5 & 1.9 & 2.4 & 3.8 & 2.3 & 5.4 & 5.3 & 3.7 & 2.3 \\
 \hline
$\ud$    &$\epsz$& $ 0.021_{-0.066}^{+0.065}$ & $ 0.090_{-0.115}^{+0.043}$ & 
                   $ 0.021_{-0.066}^{+0.065}$ & $ 0.091_{-0.118}^{+0.044}$ & 
                   $ 0.068_{-0.078}^{+0.016}$ & $ 0.000_{-0.032}^{+0.031}$ & 
                   $ 0.063_{-0.082}^{+0.031}$ & $ 0.063_{-0.083}^{+0.031}$ & 
                   $ 0.122_{-0.052}^{+0.023}$ \\
         &$\lamp$& 2.9 & 2.5 & 2.9 & 2.5 & 3.2 & 4.6 & 3.0 & 3.0 & 2.5 \\
         &$\lamm$& 3.6 & 4.3 & 3.6 & 4.3 & 6.6 & 4.6 & 5.0 & 5.0 & 5.4 \\
 \hline
\end{tabular}\end{center}}
\caption[foo]{\label{tab:ccres}
  Results of the contact interaction fits to the angular distributions for
  non-radiative $\eetoee$, $\eetomumu$, $\eetotautau$ and the cross-sections
  for $\eetoqq$. Results at centre-of-mass energies of 
  130--183~GeV~\cite{bib:OPAL-SM172,bib:OPAL-SM183} are also included.
  The combined results include all leptonic angular distributions and the 
  hadronic cross-sections. The numbers in square brackets are the values of
  [$\eta_{\mathrm{LL}}$,$\eta_{\mathrm{RR}}$,$\eta_{\mathrm{LR}}$,
  $\!\eta_{\mathrm{RL}}$] which define the models.
  Note that the values of $\eta$ for the $\overline{\cal{O}}_{\mathrm{DB}}$ 
  model~\cite{bib:CInew2} have been scaled down by a factor of 4
  compared with those used in~\cite{bib:OPAL-SM183}.  
  $\epsz$ is the fitted value of $\varepsilon = 1/\Lambda^{2}$,
  $\Lambda_{\pm}$ are the 95\% confidence level limits.
  The units of $\Lambda$ are TeV, those of $\epsz$ are $\mathrm{TeV}^{-2}$.
}
\end{minipage}
\end{sideways}
\end{table}

\begin{table}[htb]
\centering
\begin{tabular}{|c|c|c||c|}
\hline
 Processes   & $\sqrt{s}$ [GeV]  &$\lambda$  &$M_{\rm s}$ [TeV] (95\% C.L.) \\
\hline
\hline
$\epem \rightarrow \mumu$  &  183,189 & $+1$ &  0.60   \\
\cline{3-4}
                &          & $-1$ &  0.63   \\
\hline
$\epem \rightarrow \tautau$ &  183,189 & $+1$ &  0.63   \\
\cline{3-4}
                &          & $-1$ &  0.50   \\
\hline
\hline
$\epem \rightarrow \mumu, \tautau$ &  183,189 & $+1$ &  0.68   \\
\cline{3-4}
                &          & $-1$ &  0.61   \\
\hline
\hline
\end{tabular}
\caption[]{Lower limits on $M_{\rm s}$ at 95\% confidence level.
 }
\label{tab:grav}
\end{table}

%
\clearpage
%
%
%
%
%
%
%
%
%
%
%
%
%

\begin{thebibliography}{99}

\bibitem{bib:OPAL-SM183}
  OPAL Collab., G.~Abbiendi et~al., 
  \EPJ\ {\bf C6} (1999) 1.

\bibitem{bib:OPAL-SM172}
  OPAL Collab., K.~Ackerstaff et~al., 
  \EPJ\ {\bf C2} (1998) 441.

\bibitem{bib:ADL-SM}
  \ALEPHColl, D.~Buskulic et~al., \PhysLett\ {\bf B378} (1996) 373; \\
  \ALEPHColl, R.~Barate et~al., {\sl Study of Fermion Pair Production
   in \epem\ Collisions at 130--183~GeV}, CERN-EP/99-042, March 1999,
   submitted to \EPJ\ C; \\
  \DELPHIColl, P.~Abreu et~al., {\sl Measurement and Interpretation of 
   Fermion-Pair Production at LEP Energies from 130 to 172 GeV}, 
   CERN-EP/99-05, January 1999, to be published in \EPJ\ C; \\
  \LthreeColl, M.~Acciarri et~al., \PhysLett\ {\bf B370} (1996) 195; \\
  \LthreeColl, M.~Acciarri et~al., \PhysLett\ {\bf B407} (1997) 361; \\
 L3 Collab., M.~Acciarri et al., \PhysLett\ {\bf B414} (1997) 373; \\
 L3 Collab., M.~Acciarri et al., \PhysLett\ {\bf B433} (1998) 163.

\bibitem{bib:ADD}
 N.~Arkani-Hamed, S.~Dimopoulos and G.~Dvali,
 \PhysLett\  {\bf B429} (1998) 263; \\
 I.~Antoniadis, N.~Arkani-Hamed, S.~Dimopoulos and G.~Dvali,
 \PhysLett\  {\bf B436} (1998) 257; \\
 N.~Arkani-Hamed, S.~Dimopoulos and G.~Dvali, \PhysRev\ {\bf D59}
 (1999) 86004.

\bibitem{bib:Giudice}
 G.F.~Giudice, R.~Rattazzi and J.D.~Wells, \NPhys\ {\bf B544} (1999) 3.

\bibitem{bib:OPAL-detector}
  \OPALColl, K.~Ahmet et~al., \NIM\ {\bf A305} (1991) 275.

\bibitem{bib:OPAL-SI}
  S.~Anderson et~al., \NIM\ {\bf A403} (1998) 326.

\bibitem{bib:OPAL-SW}
  B.E.~Anderson et~al., \IEEENS\ {\bf 41} (1994) 845.

\bibitem{bib:OPAL-TR}
  M.~Arignon et~al., \NIM\ {\bf 313} (1992) 103; \\
  M.~Arignon et~al., \NIM\ {\bf 333} (1993) 330.

\bibitem{bib:OPAL-DAQ}
  J.T.~Baines et~al., \NIM\ {\bf A325} (1993) 271; \\
  D.G.~Charlton, F.~Meijers, T.J.~Smith, P.S.~Wells, \NIM\ {\bf A325} 
  (1993) 129.

\bibitem{bib:ELEP}
  The LEP Energy Working Group, {\sl Evaluation of the LEP centre-of-mass
  energy for data taken in 1998}, LEP Energy Working Group 99/01, 
  March 1999; \\
  LEP Energy Working Group, A~Blondel et al., CERN-EP/98-191,
  CERN-SL/98-073, submitted to \EPJ\ C.

\bibitem{bib:gopal}
  J.~Allison et~al., \NIM\ {\bf A317} (1992) 47.

\bibitem{bib:bhlumi}
  S.~Jadach et~al., \CPC\ {\bf 102} (1997) 229.

\bibitem{bib:bhlumi_err}
  W.~Placzek, S.~Jadach, M.~Melles, B.F.L.~Ward and S.A.Jost,
  CERN-TH/99-07.

\bibitem{bib:zfitter}
  D.~Bardin et~al., DESY 99-070; \\
  D.~Bardin et~al., CERN-TH 6443/92; \\
  D.~Bardin et~al., \PhysLett\ {\bf B255} (1991) 290; \\
  D.~Bardin et~al., \NPhys\ {\bf B351} (1991) 1; \\
  D.~Bardin et~al., \ZPhys\ {\bf C44} (1989) 493.\\
  We use ZFITTER version 6.10 with default parameters, except 
  {\tt INTF}=0,  {\tt FINR}=0 and {\tt ISPP}=--1,
  and with the following input parameters: $\mPZ$=91.1863~GeV,
  $\mtop$=175~GeV, $\mHiggs$=175~GeV, 
  $\alphas(\mPZ)$=0.118.
  
\bibitem{bib:alibaba}
  W.~Beenakker~et~al., \NPhys\ {\bf B349} (1991) 323.

\bibitem{bib:OPAL-LS91}
  \OPALColl, P.D.~Acton et~al., \ZPhys\ {\bf C58} (1993) 219.

\bibitem{bib:OPAL-LS92}
  \OPALColl, R.~Akers et~al., \ZPhys\ {\bf C61} (1994) 19.

\bibitem{bib:OPAL-LS90}
  \OPALColl, G.~Alexander et~al., \ZPhys\ {\bf C52} (1991) 175.

\bibitem{bib:OPAL-WW183} 
  OPAL Collab., G.~Abbiendi et~al., \EPJ\ {\bf C8} (1999) 191.

\bibitem{bib:pythia}
  T.~Sj\"ostrand, \CPC\ {\bf 82} (1994) 74.

\bibitem{bib:KK2f}
  S.~Jadach, B.F.L.~Ward and Z.~W\c{a}s, \PhysLett\ {\bf B449} (1999) 97.
  
\bibitem{bib:herwig}
 G.~Marchesini et~al., \CPC\ {\bf 67} (1992) 465.

\bibitem{bib:phojet}
  R.~Engel and J.~Ranft, \PhysRev\ {\bf D54} (1996) 4244.

\bibitem{bib:twogen}
  A.~Buijs et al., \CPC\ {\bf 79} (1994) 523.

\bibitem{bib:OPAL-f2gam}
 \OPALColl\, K.~Ackerstaff et~al., \ZPhys\ {\bf C74} (1997) 33.

\bibitem{bib:grc4f}
  J.~Fujimoto et~al., \CPC\ {\bf 100} (1997) 128.

\bibitem{bib:excalibur}
  F.A.~Berends, R.~Pittau and R.~Kleiss, \CPC\ {\bf 85} (1995) 437.

\bibitem{bib:OPAL-gg189}
  OPAL Collab., G.~Abbiendi et~al., 
  {\sl Multi-photon production in \epem\ collisions at $\sqrt{s}$ = 189~GeV},
  CERN-EP/99-088, submitted to \PhysLett\ B.

\bibitem{bib:mH_exp}
  The LEP Working Group for Higgs Boson Searches, ALEPH, DELPHI, L3
  and OPAL Collaborations, {\sl Limits on Higgs Boson Masses from Combining
  the Data of the Four LEP Experiments at Energies up to 183~GeV},
  CERN-EP/99-060.

\bibitem{bib:mH_ew}
  The LEP Electroweak Working Group and the SLD Heavy Flavour and Electroweak
  Groups, ALEPH, DELPHI, L3, OPAL and SLD Collaborations,
  {\sl A Combination of Preliminary Electroweak Measurements and
  Constraints on the Standard Model}, CERN-EP/99-015.

\bibitem{bib:topaz0}
 G.~Montagna, O.~Nicrosini, G.~Passarino, F.~Piccinini and R.Pittau,
 \CPC\ {\bf 76} (1993) 328; \\
 G.~Montagna, O.~Nicrosini, G.~Passarino and F.~Piccinini,
 \CPC\ {\bf 93} (1996) 120; \\
 G.~Montagna, O.~Nicrosini, F.~Piccinini and G.~Passarino,
 hep-ph/9804211. \\
 We use TOPAZ0 version 4.4 with default parameters, except 
 {\tt ONP='Y'}, and with the same input parameters as for ZFITTER.
 
\bibitem{bib:bhwide}
  S.~Jadach, W.~Placzek, B.F.L.~Ward, \PhysLett\ {\bf B390} (1997) 298.

\bibitem{bib:gentle}
 D.~Bardin et~al., \NPhys\  Proc. Suppl. {\bf 37B} (1994) 148.

\bibitem{bib:fermisv}
  J.~Hilgart, R.~Kleiss, F.~Le Diberder, Comp.~Phys.~Comm. {\bf 75} (1993)
 191.

\bibitem{bib:rdata}
HRS Collab., D.~Bender et~al., \PhysRev\ {\bf D31} (1985) 1; \\
MAC Collab., E.~Fernandez et~al., \PhysRev\ {\bf D31} (1985) 1537; \\
PLUTO Collab., C.~Berger et~al., \PhysLett\ {\bf B81} (1979) 410; \\
CELLO Collab., H.J.~Behrend et~al., \PhysLett\ {\bf B183} (1987) 400; \\
JADE Collab., W.~Bartel et~al., \PhysLett\ {\bf B129} (1983) 145; \\
JADE Collab., W.~Bartel et~al., \PhysLett\ {\bf B160} (1985) 337; \\
MARKJ Collab., B.~Adeva et~al., \PRL\ {\bf 50} (1983) 799; \\
MARKJ Collab., B.~Adeva et~al., \PhysRev\ {\bf D34} (1986) 681; \\
TASSO Collab., R.~Brandelik et~al., \PhysLett\ {\bf B113} (1982) 499; \\
TASSO Collab., M.~Althoff et~al., \PhysLett\ {\bf B138} (1984) 441; \\
AMY Collab., T.~Mori et~al., \PhysLett\ {\bf B218} (1989) 499; \\
TOPAZ Collab., I.~Adachi et~al., \PRL\ {\bf 60} (1988) 97; \\
VENUS Collab., H.~Yoshida et~al., \PhysLett\ {\bf B198} (1987) 570; \\
TOPAZ Collab., K.~Miyabayashi et~al., \PhysLett\ {\bf B347} (1995) 171.

\bibitem{bib:alrun}
 TOPAZ Collab., I.~Levine et~al., \PRL\ {\bf 78} (1997) 424. 

\bibitem{bib:MK_alphaem}
 M.~Kobel, {\sl Direct Measurements of the Electromagnetic Coupling
 Constant at Large $q^2$}, FREIBURG-EHEP 97-13,
 Contributed paper to the XVIII International Symposium on Lepton
 Photon Interactions, Hamburg, July 1997.

\bibitem{bib:CAGE}
 M.E.~Cage et al., IEEE Trans.~Instrum.~Meth.~{\bf 38} (1989) 284.

\bibitem{bib:Eichten}
  E.~Eichten, K.~Lane, M.~Peskin, \PRL\ {\bf 50} (1983) 811.

\bibitem{bib:contacttable}
 {See for example {\sl Contact interactions and 
  new heavy bosons at HERA: a model independent analysis}, 
  P. Haberl, F. Schrempp, H.U. Martyn, 
  in {Proceedings, Physics at HERA}, {\bf vol. 2}, (1991) 1133. }

\bibitem{bib:OPAL-CI} 
  \OPALColl, G.~Alexander et~al., \PhysLett\ {\bf B387} (1996) 432.

\bibitem{bib:CInew2}
  G.J.~Gounaris, D.T.~Papadamou, F.M.~Renard, 
  \PhysRev\ {\bf D56} (1997) 3970.

\bibitem{bib:CI-CDF}
  CDF Collab., F.~Abe et al., \PRL\ {\bf 79} (1997) 2198.

\bibitem{bib:CI-H1}
  H1 Collab., S.~Aid et al., \PhysLett\ {\bf B353} (1995) 578.

\bibitem{bib:CI-atomic}
  A.~Deandrea, \PhysLett\ {\bf B409} (1997) 277.

\bibitem{bib:rpvsup}   
  J.~Wess,  J.~Bagger,    {\it Supersymmetry  and
    Supergravity}  (Princeton  University Press, 1983);  \\  
  H.P.~Nilles, \PhysRep\ {\bf 110}   (1984) 1; \\ 
  H.E.~Haber,  G.E.~Kane,  \PhysRep\ {\bf 117} (1985) 75;  \\
  R.~Barbieri, Riv.~Nuovo~Cim. {\bf 11} (1988) 1; \\
  P.~West, {\it Introduction to  Supersymmetry and Supergravity} (World
  Scientific, 1986). 

\bibitem{bib:rpvsnu}
  J.~Kalinowski, R.~R\"{u}ckl, H.~Spiesberger, P.M.~Zerwas,
    \PhysLett\ {\bf B406} (1997) 314.

\bibitem{bib:ALEPH-rpv}
 \ALEPHColl, R.~Barate et al., \EPJ\ {\bf C4} (1998) 433; \\
 \OPALColl, G.~Abbiendi et al., {\sl Search for R-parity Violating Decays
  of Scalar Fermions at LEP}, CERN-EP/99-043, submitted to \EPJ\ C.

\bibitem{bib:Hewett}
 J.L.~Hewett, \PRL\ {\bf 82} (1999) 4765.

\bibitem{bib:Peskin}
 E.A.~Mirabelli, M.~Perelstein, M.E.~Peskin, \PRL\ {\bf 82} (1999) 2236.

\bibitem{bib:Rizzo}
 T.G.~Rizzo, \PhysRev\ {\bf D59} (1999) 115010.

\bibitem{bib:BK}
 F.A.~Berends and R.~Kleiss, Nucl. Phys. {\bf B186} (1981) 22.

\end{thebibliography}
\end{document}